\documentclass[12pt]{revtex4}

\usepackage{here}
\usepackage{graphicx}
\usepackage{amsmath}
\usepackage{amssymb}
\usepackage{color}
\usepackage{xspace}
\usepackage[latin1]{inputenc}
\usepackage[T1]{fontenc}
\usepackage{hyperref}

\newcommand{\be}{\begin{displaymath}}
\newcommand{\ee}{\end{displaymath}}
\newcommand{\bn}{\begin{equation}}
\newcommand{\en}{\end{equation}}

\newcommand{\red}[1]{\textcolor{black}{ #1}}
\newcommand{\blue}[1]{\textcolor{black}{ #1}}

\newcommand{\gyro}{{\sc gyro}\xspace}
\newcommand{\new}[1]{\textcolor{black}{#1}}
\newcommand{\newnew}[1]{\textcolor{black}{#1}}
\newcommand{\newred}[1]{\textcolor{black}{ #1}}
\begin{document}

\title[Core micro-instability analysis of JET hybrid and baseline
    discharges with carbon wall]{Core micro-instability analysis of
  JET \red{hybrid and baseline discharges with carbon wall}}

\author{S. Moradi$^{1,2}$, I. Pusztai$^{2,3}$,
  I. Voitsekhovitch$^{4}$, L. Garzotti$^{4}$, C. Bourdelle$^5$,
  M. J. Pueschel$^{6}$,  I. Lupelli$^{4}$, M. Romanelli$^{4}$, and the JET-EFDA
  contributors \footnote{F. Romanelli et al. {\em Proc. 24th Int. Conf. on Fusion
  Energy} (2012) (San Diego, USA, 2012) paper OV/1-3.
  www-naweb.iaea.org/napc/physics/FEC/FEC2012/index.htm.}}
  \address{$^{1}$ Ecole Polytechnique, CNRS UMR7648, LPP, F-91128, Palaiseau, France\\ 
  $^{2}$ Applied Physics,
  Chalmers University of Technology,
  G\"{o}teborg 41296, Sweden\\ 
  $^{3}$ Plasma Science and Fusion
  Center, Massachusetts Institute of Technology, Cambridge MA, 02139,
  USA \\
  $^{4}$ JET-EFDA, Culham Science Centre, Abingdon, OX14 3DB, UK \\
  $^5$ IRFM, CEA, F-13108 Saint Paul-Lez-Durance, France \\ 
  $^{6}$ University of Wisconsin-Madison, Madison,
  Wisconsin 53706, USA}

\begin{abstract}
The core micro-instability characteristics of hybrid and baseline
plasmas in a selected set of JET plasmas with carbon wall are
investigated through local linear and non-linear \new{and global
  linear} gyro-kinetic simulations with the \gyro code [J. Candy and
  E. Belli, General Atomics Report GA-A26818 (2011)]. In particular,
we study the role of plasma pressure on the micro-instabilities, and
scan the parameter space for the important plasma parameters
responsible for the onset and stabilization of the modes under
experimental conditions. We find that a good core confinement due to
strong stabilization of the micro-turbulence driven transport can be
expected in the hybrid plasmas due to the stabilizing effect of the
fast ion pressure that is more effective at the low magnetic shear of
the hybrid discharges. While parallel velocity gradient
destabilization is important for the inner core, at outer radii the
hybrid plasmas may benefit from a strong quench of the turbulence
transport by $\mathbf{E}\times\mathbf{B}$ rotation shear.
    \end{abstract}

\maketitle

\section{Introduction}
In recent years there has been an increasing worldwide effort in the
development of the so-called hybrid or improved H-mode scenarios as a
hybrid between an Advanced Tokamak and a baseline plasma
\cite{shimada,mcdonald2008,Gormezano2004,Gormezano2007}. In these
scenarios, by optimizing the current density profile, an enhanced
normalized confinement can be achieved, as compared to the ITER
baseline scenario, the ELMy H-mode.


Also, a higher bootstrap fraction and a lower requirement for induced
current allows longer pulses in the hybrid mode. A limiting factor for
the plasma performance and achievable normalized plasma pressure
\red{$\beta_N(=\bar{\beta} a[{\rm m}] B_T[{\rm T}]/I_P[{\rm MA}]$,
  with $\bar{\beta}$,} the volume averaged normalized pressure, $a$
the plasma minor radius, $B_T$ the toroidal magnetic field and $I_P$
the plasma current), is the onset of core MHD activity such as
$m/n=2/1, 3/2$ neoclassical tearing modes \cite{Buttery2003}. However,
by tailoring the current profile the destabilization of these large
MHD activities can be avoided by slowing down the current diffusion
time and hence prolonging the time period when the safety-factor
profile is flat but kept above 1 ($q_0>1$) and a radially extended low
magnetic shear region is present in the plasma core. These
developments gave rise to the so-called hybrid or improved H-mode
scenarios \cite{Gormezano2007}.

The method used at Joint European Torus (JET) for slowing down the
current diffusion in the hybrid scenarios is through an overshoot of
the current in the pre-heating phase just before reducing it to its
flat-top value during the main heating phase
\cite{joffrin2005,joffrin2010,Hobirk2012}. This method allows for the
broadening of the current profile with flat core $q$-profile over a
large part of plasma radius which results in an enhancement of the
confinement factor $H_{98}(y,2)$ compared to what could be obtained
with a regular ramp-up scenario (baseline scenario). For the
definition of $H_{98}(y,2)$ see Ref. \cite{iter} and references
therein.

During the 2008-2009 experimental campaigns at the JET, a remarkable
improvement in the normalized confinement was achieved in hybrid
scenarios ($H_{98}(y,2)\sim1.3-1.4$) in both high ($\delta =0.4$) and
low ($\delta=0.2$) triangularity plasmas
\cite{joffrin2010,Beurskens2013}. The plasma configurations are
optimized to reduce the wall interactions and therefore the recycling
of the wall neutrals. In the hybrid plasmas lower density and higher
temperature are achieved.



In a recent work, Ref. \cite{Beurskens2013}, the
confinement properties of the hybrid and baseline plasmas in a
database of 112 discharges in JET with Carbon Fiber Composite (CFC)
wall have been analyzed. Here, it has been shown that the confinement
factor $H_{98}(y,2)$ in JET hybrid plasmas are typically $1 <
H_{98}(y,2)< 1.5$ while in the baseline ELMy H-mode plasmas typical
values are $H_{98}(y,2)\sim 1$. However, the underlying physics basis
for the observed increased normalized confinement remain somewhat
unclear, making the hybrid plasmas an interesting choice for modeling
and trying to explain the underlying mechanisms responsible for their
deviation from the ``known'' H-mode confinement. Moreover, due to
their high normalized pressure, electromagnetic effects and their
impact on the turbulent transport driven by micro-instabilities has to
be investigated.
  
Recent reports have shown the significant role of electromagnetic
modes such as Micro-Tearing Modes (MTMs) and Kinetic-Ballooning Modes
(KBMs) on the heat and particle transport in the core of fusion
plasmas \cite{smoradipop2013,smoradipop2012,Pueschel08,Pueschel10,
  Ishizawa2013,GuttenfelderPRL2011,DoerkPRL2011}. In the presence of
high $\beta(=8\pi\langle p \rangle/B_T^{2}$, with $\langle p \rangle$
being the volume average kinetic pressure), it has been shown that the
electrostatic modes such as Ion Temperature Gradient Modes (ITG) may
be fully stabilized while electromagnetic modes are destabilized as
$\beta$ is increased and therefore can give rise to a new regime with
high turbulent transport. However, since in the hybrid plasmas the
confinement shows improvement despite the increased $\beta$ it is
important to understand the possible stabilization mechanisms for the
electromagnetic modes which are at play in these plasmas. Furthermore,
the impact of plasma shape \red{on} electromagnetic modes is shown to
be stronger than \red{on} electrostatic modes
\cite{smoradipop2013,smoradipop2012,Pueschel08}\red{.} Therefore, it
is important to include plasma shaping when studying these modes.

In this paper therefore, we investigate the \red{characteristics of
  core micro-instabilities in} two hybrid and two baseline plasmas in
JET selected from the database analyzed in Ref. \cite{Beurskens2013}
through local linear and non-linear \new{and global linear} gyro-kinetic simulations
\red{using} the \gyro code \cite{gyro,GYRO}. In particular, we study
the role of plasma pressure micro-instabilities, and scan the
parameter space for the important plasma parameters responsible for
the onset and stabilization of the modes under realistic conditions. 

\newred{Even considering the global profile variations, we find the
  possibility of linearly unstable KBMs in the core in a parameter
  regime representative of hybrid discharges, which is the result of
  an extended region with very small magnetic shear.  Our findings
  suggest that a strong stabilization of the micro-turbulence in the
  core needed to explain the good inner core confinement in the hybrid
  plasmas can be a result of stabilization by the fast ion pressure
  which is more effective at low magnetic shear. We also find that
  parallel velocity gradient has a non-negligible destabilizing
  contribution in the inner core in some of the studied discharges,
  while for the outer core radii the hybrid plasmas may benefit from a
  strong reduction of turbulent transport by
  $\mathbf{E}\times\mathbf{B}$ rotational shear.}

The remainder of the paper is organized as follows. In
Sec.~\ref{sec:input} \red{we discuss the studied discharges, the data
  analysis, and the set of input parameters used in the
  simulations. In Sec.~\ref{sec:1} we present the gyrokinetic
  microinstability analysis of the chosen scenarios. To assess the
  sensitivity of the results and to identify possible stabilizing
  mechanisms, we perform scans in plasma profile and geometry
  parameters around their nominal values with special emphasis on
  finite-$\beta$, plasma shaping, and fast ion effects. \blue{The
    effect of rotation is addressed and nonlinear simulation results
    are given in Sec.~\ref{sec:nonlin}.} Finally, we discuss or
  results and conclude in Sec.~\ref{sec:conclusions}.}

\section{Experimental data used in simulations}
\label{sec:input}
Four JET discharges are chosen for our study: two hybrid discharges:
77922 and 75225 with high and low triangularity respectively, and two
baseline discharges: 76679 and 78682, again with high and low
triangularity respectively. These discharges have been chosen from the
database presented in Ref.\cite{Beurskens2013} as part of the
2008-2009 JET experimental campaigns with CFC wall. More detailed
descriptions of the experimental set up in these discharges can be
found in Refs.~\cite{Beurskens2013,Hobirk2012}. \newred{Here, we would like to
note that these shots are arbitrary selected examples of hybrid
and baseline plasmas, thus, the conclusions drawn for
these discharges do not necessarily extrapolate to the whole 
group of hybrid and baseline H-mode discharges.}

\begin{figure}[htbp]
\begin{center}
 \includegraphics[width=0.8\textwidth]{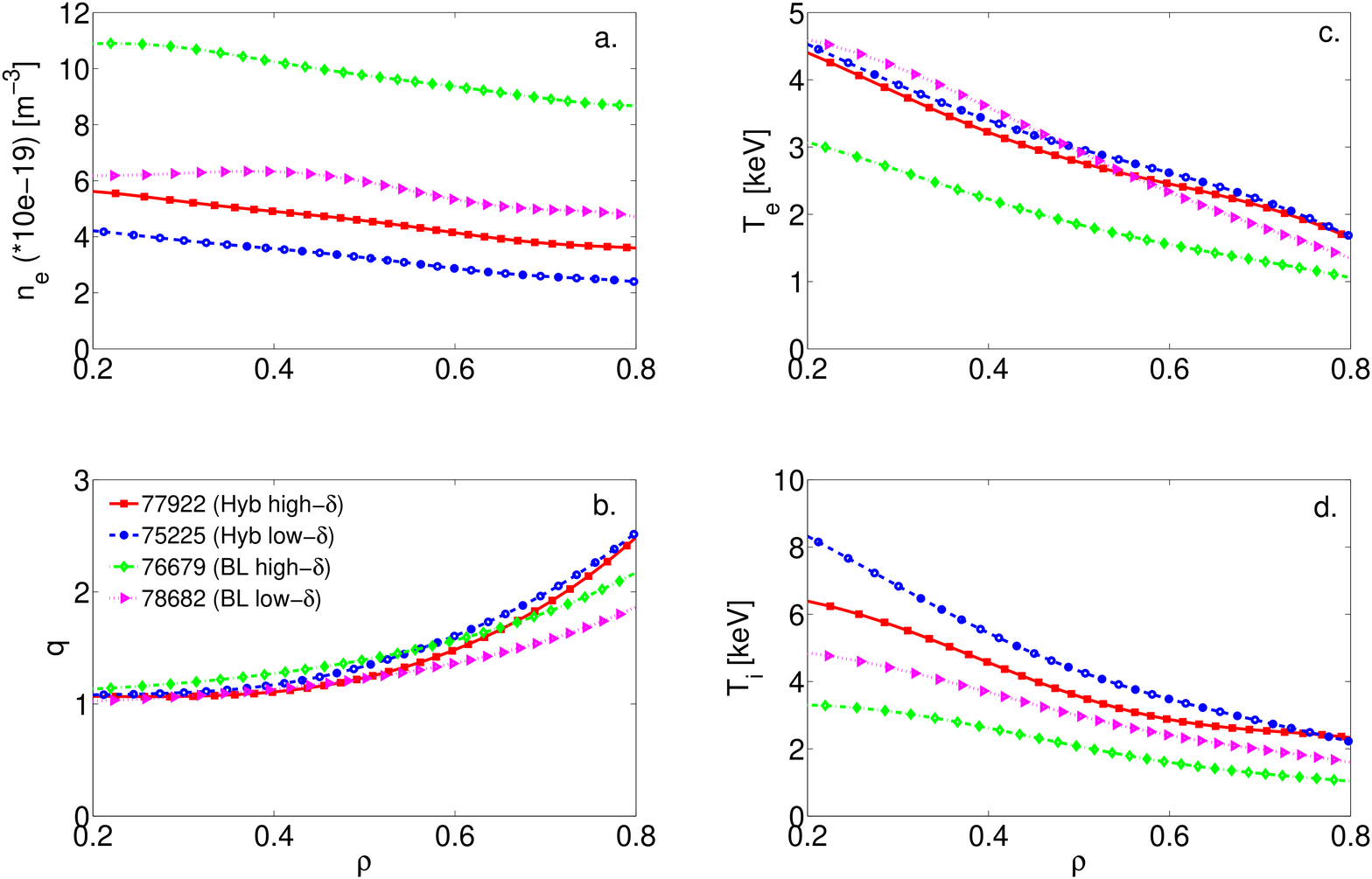}\\ \includegraphics[width=0.4\textwidth]{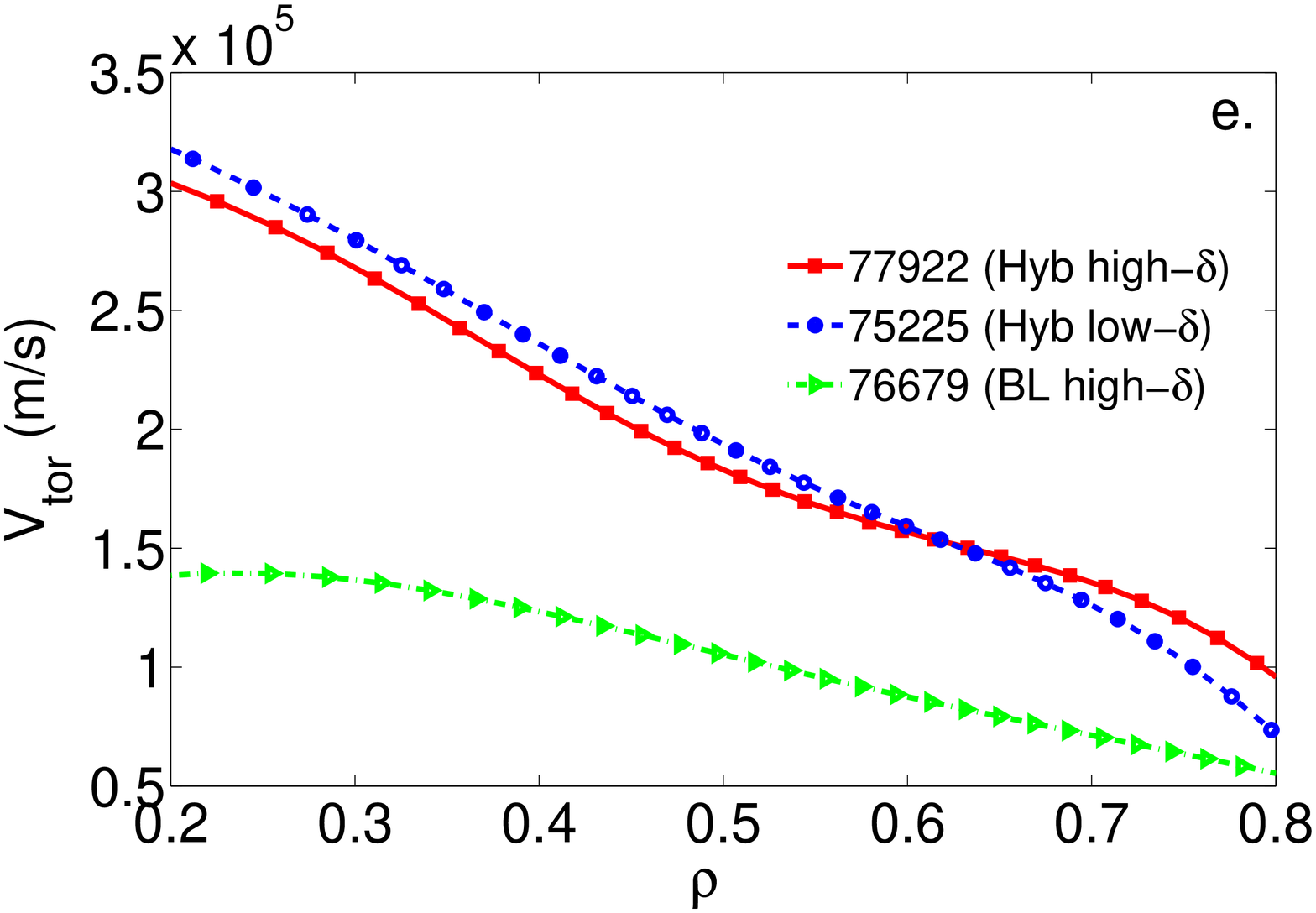}
\caption{Input profiles: (a) electron density, \red{(b) safety factor,
    (c) electron temperature,} (d) ion temperature, and (e) toroidal
  rotation velocity calculated by TRANSP code. Solid (red) lines show
  the profiles for JET discharge 77922, dashed (blue) lines for 75225,
  dashed-dotted (green) lines for 76679, and dotted (mauve) lines for
  78682 discharges.}
\label{fig1}
\end{center}
\end{figure}

Figure \ref{fig1} shows the profiles of \red{electron} plasma density
$n_e$, ion and electron temperatures $T_{i,e}$, toroidal rotation
speed $V_{tor}$ (for the baseline, only \red{in} the high-$\delta$
case) measured at outboard mid-plane and re-mapped on TRANSP
\cite{transp} equilibrium, and safety factor $q$ for the selected
discharges. In hybrid discharges the $q$-profiles have been
reconstructed in the EFIT simulations constrained by Motional Stark
Effect (MSE) measurements \cite{Lao1990,Hobirk2012}. In baseline
discharges the $q$-profiles have been simulated by TRANSP using NCLASS
module for current conductivity and bootstrap current. The q-profile
at the beginning of NBI heating phase reconstructed with EFIT
constrained by magnetic probe measurements has been used as \red{an}
initial condition for current diffusion simulations. However, the
calculated $q$ in the core may be too low in these cases.

\begin{figure}[htbp]
\begin{center}
 \includegraphics[width=0.8\textwidth]{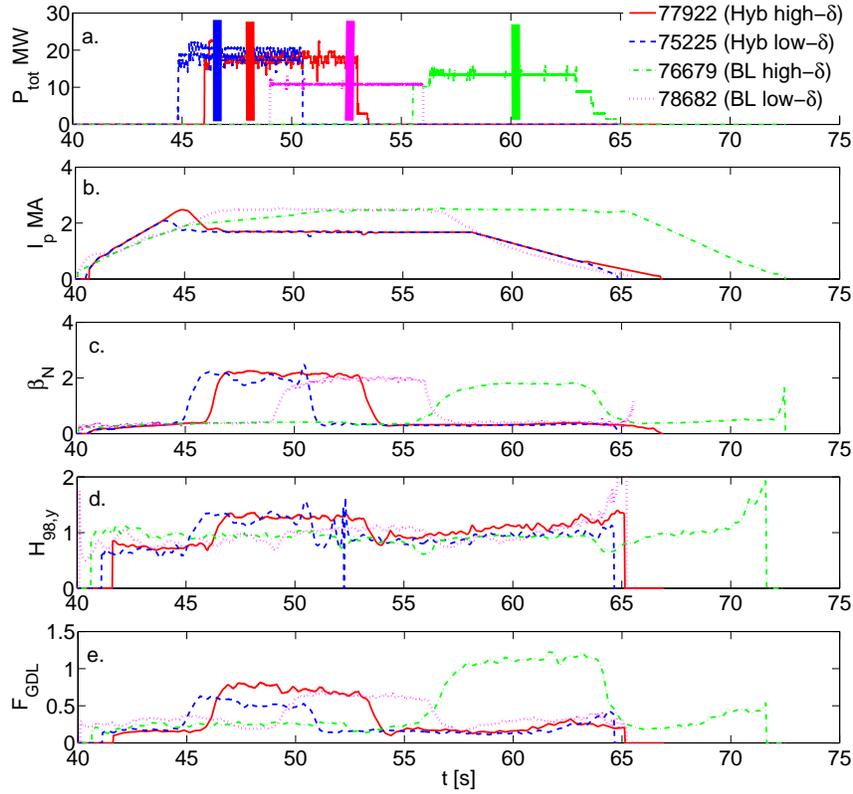} 
\caption{Time traces of main plasma parameters: (a) total input NBI
  power $P_{tot}$, (b) plasma current $I_P$, (c) normalized thermal
  pressure $\beta_N$, (d) the global confinement enhancement factor
  $H_{98}(y,2)$ and (e) the Greenwald density fraction
  $F_{GDL}$. Solid (red) lines show the values for JET discharge
  77922, dashed (blue) lines for 75225, dashed-dotted (green) lines
  for 76679, and dotted (mauve) lines for 78682 discharges. The
  highlighted bars in (a) denote the time \red{periods} for which the
  profiles were averaged over, and are color coded similarly to their
  corresponding curves.}
\label{fig2}
\end{center}
\end{figure}

The plasma profiles shown in Fig. \ref{fig1} are averaged over a time
window of $0.5\,\rm{s}$ over the period of the highest $H_{98}(y,2)$
factor with nearly stationary temperatures and density. The hybrid
plasmas do not show any Neoclassical Tearing \red{Mode (NTM)} activity
during the selected time window, while in baseline plasmas an $n=1$ mode
is present. Figure \ref{fig2} illustrates time traces of the plasma
parameters: total input NBI power $P_{tot}$, plasma current $I_P$,
normalized \red{thermal pressure}, the global confinement enhancement
factor $H_{98}(y,2)$ and the Greenwald density fraction $F_{GDL}\sim
\bar{n}/n_{G}$ are shown (where $\bar{n}$ is the line averaged
density, and $n_{G}[\rm{ m^{-3}}]=10^{20} I_{p}[{\rm MA}]/\pi a[{\rm
    m}]^2$ with $a$ being the minor radius) and the highlighted bars
denote the time period over which the profiles were averaged. The
characteristics of the hybrid scenarios with the current overshoot in
the preheating phase as compared to baseline regular ramp-up scenarios
can be seen on the time traces of the plasma current in
Fig. \ref{fig2}(b). Also, note that hybrids operate at lower plasma
currents than baselines.

As seen in Fig. \ref{fig1}(a), plasma density in hybrid plasmas is lower
compared to those of the baselines as a result of lower current. The
fueling level for the baseline discharges is: \red{$\Gamma_{D2} \sim
  0.4-4 \times 10^{22}\; {\rm electron\; s^{-1}}$}, whereas for the
hybrid plasmas the fueling is generally low, with \red{$\Gamma_{D2}
  \sim 0-0.25 \times 10^{22} \;{\rm electron\; s^{-1}}$}. In the
hybrid plasmas it is difficult to increase the density by higher gas
fueling as it can result in degradation of pedestal
confinement\red{, leading to a} reduction of density.  The fueling
waveforms for the two selected hybrid discharges can be found in
Ref.\cite{Beurskens2013} figure 4.

The input powers were also different between hybrid \red{($P_{tot}\sim
  17-20\,{\rm MW}$)} and baseline plasmas \red{($P_{tot}\sim
  10-13\,{\rm MW}$)}, leading to the higher temperatures and toroidal
rotation in hybrids as seen in Fig. \ref{fig1}(c,d) and (e). No ICRH
heating is applied in these hybrid plasmas, but to avoid impurity
accumulation a small amount of central ICRH heating ($P_{ICRH} <
10\%P_{NBI}$) has been applied in baseline plasmas
\cite{Beurskens2013}. As shown in Fig. \ref{fig1}(b) the safety factor
in hybrid discharges has a broader flat region in the core and it is
higher at the edge, as compared to the baseline discharges.

For our core micro-instability analysis two radial positions are
chosen: $r/a=0.3$ and $0.6$, where $r$ is the half-width of the flux
surfaces, and $a$ is $r$ at the separatrix \red{(as defined in
  \cite{GYRO})}. Using the profiles shown in Fig. \ref{fig1}, tables
\ref{table1}-\ref{table4} present the local plasma parameters used in
our study for each of the four selected discharges at the two radial
positions. Magnetic geometry parameters, such as those describing the
shaping of flux surfaces or the $q$-profile, can be changed
independently in the \gyro code, see Refs.\cite{waltz09,candy}. Here
in the hybrid cases these parameters are based on magnetic geometries
  from EFIT constrained by MSE measurements, whereas in baselines they are
  simulated by TRANSP using NCLASS module for current conductivity and
  bootstrap current. The density and temperature scale lengths are
  defined as: $L_{n}=-[\partial (\ln{n})/\partial r]^{-1}$,
  $L_{T}=-[\partial (\ln{T})/\partial r]^{-1}$. The safety factor is
  $q=d\chi_t/d \psi$, with \red{$2\pi\chi_t$} is the toroidal flux, and $2\pi
  \psi$ being the poloidal flux, and the magnetic shear is
  $s=(r/q)dq/dr$. In these tables, $\beta_e$ is calculated following
  the expression:
\begin{equation}
\beta_{e}=\frac{8 \pi ( n_e[10^{19}/m^3]\;10^{-6} \;10^{19} )(
  T_e[keV]\;1.6022\;10^{-9} )}{(10^{4}\; B_{unit}[T] )^2},
\label{betae}
\end{equation}
where $B_{unit}$ is defined as the effective field strength, see
Refs. \cite{waltz09,candy09},
\begin{equation}
B_{unit}=\frac{1}{r}\frac{d\chi_t}{dr}.
\label{bunit}
\end{equation}
Note that for a pure hydrogenic plasma with $T_e=T_i$, we have
$\beta=2 \beta_e$.

Furthermore, $\epsilon$ is the inverse aspect ratio, $\kappa$ is the
elongation, $\delta$ is the triangularity, and
\red{$s_{\delta}=r\partial\delta/\partial r$}, and
$s_{\kappa}=(r/\kappa)\partial\kappa/\partial r $ are the
triangularity and elevation shear, respectively. The convention of
defining $\beta_e$ in terms of $B_{unit}$ is specific to {\sc gyro},
and not to gyrokinetic codes in general. We note that $B_{unit}$ in
shaped plasmas is not equal to the on-axis magnetic field $\bar{B}$,
because the effective radius $r$ is defined as the half width of the
flux surface at the elevation of its centroid and thus the toroidal
flux inside a given flux surface is generally not equal to $r^2\pi
\bar{B}$. Since for elongated flux surfaces ($\kappa>1$) the area of
the cross section of the flux surface is larger than $\pi r^2$, the
effective field is roughly $\kappa$ times larger than the on-axis
field. It is important that it is $\beta_e$, defined with the
\emph{effective field}, that matters for the stability of MTMs/KBMs,
and it is reduced by approximately a factor of $1/\kappa^2$ compared
to a $\beta_e$ when defined in terms of $\bar{B}$. This should be
taken into account when the experimental $\beta_e$ is compared to
gyro-kinetic simulations.

In the linear simulations, the normalized wave numbers $k_\theta
\rho_s$, are fixed to the given values in the tables \ref{table3} and
\ref{table4}. These wave numbers correspond to the most linearly
unstable modes with the highest growth rates. These are the baseline
cases in our study, and these parameters will be used unless otherwise
stated. The so called generalized magnetohydrodynamic $\alpha_{MHD}$
parameter is defined as (see Ref.~\cite{candy09})
\begin{equation} 
\alpha_{MHD}=-q^2 R_{0}\frac{8\pi}{B_{unit}^2}\frac{dp}{dr}c_{p}
,\label{alpha}
\end{equation}
where $R_{0}$ is the \red{major radius of the centroid of the flux
  surface}, and $p=\sum_{a}n_{a}T_{a}$ is the total plasma
pressure. $c_{p}$ is the geometric pressure gradient scaling parameter
which allows an artificial adjustment of $\alpha_{MHD}$ without
modifying the background gradients as presented in Ref.~\cite{belli}.

\begin{figure}[htbp]
\begin{center}
 \includegraphics[width=0.9\textwidth]{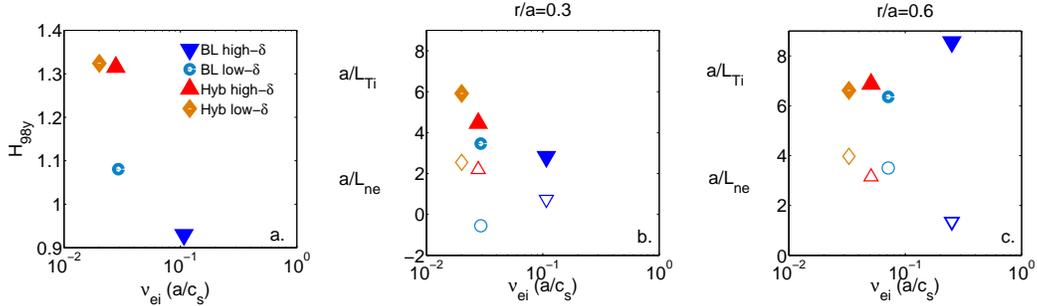}
\caption{(a) Confinement factor $H_{98}(y,2)$, and (b) input density (open
  symbols) and temperature (full symbols) gradients versus
  \red{collision frequency} $\nu_{ei}$, at inner core radius
  $r/a=0.3$, and (c) at outer core radius $r/a=0.6$. The
  down-triangles (dark blue) correspond to the values from the JET
  discharge 76679, circles (light blue) for 78682, up-triangles (red)
  for 77922, and diamonds (brown) correspond to the values from
  75225.}
\label{fig3}
\end{center}
\end{figure}

For a clearer comparison regarding the position of our selected
discharges relative to the rest of the database presented in
Ref. \cite{Beurskens2013} (see Figs. 5(b) and 14), the confinement
enhancement factor $H_{98}(y,2)$, density and temperature \red{scale
  lengths} as functions of normalized collision frequency, $\nu_{ei}
(a/c_s)$, are shown in Fig. \ref{fig3}. As seen in this figure for
both radii, the two hybrid plasmas are located at low collisionality
together with the low-$\delta$ baseline due to their lower density and
higher temperature, while the high-$\delta$ baseline is located at
high collisionality for both radii due to its high density and low
temperature.

\red{We note, that since the densities, the applied heating powers and
  total plasma currents are rather different in the cases considered
  the values of the normalized confinement factor $H_{98}(y,2)$ may
  not be indicative of the plasma confinement in absolute units; in
  fact the baseline plasmas have higher confinement times than that of
  the hybrids for in the selected discharges. Nevertheless, to identify
  the reason for the good normalized confinement in hybrids is of
  great practical importance for development of high-gain scenarios in
  tokamak reactors.}

Figures \ref{fig3}(b) and (c) 
show that the gradients of the electron density and ion temperature 
are highest in hybrids at the inner radius ($r/a=0.3$). At the outer
radius however, the high-$\delta$ baseline plasma shows the highest values. The
low-$\delta$ baseline in comparison to both hybrid plasmas, also shows
lower values for electron density and ion temperature gradients at the
inner core radius, but at the outer radius these values are comparable
to those of the hybrids. 
In the following sections we will investigate the stability
of the selected plasmas to electrostatic and electromagnetic
micro-instabilities and discuss the results and implications for the
developments of the future operational scenarios.

\begin{table}[ht]
\caption{Input parameters for plasma shape and magnetic geometry at $r/a=0.3$.}
\centering
\begin{tabular}{|c|c|c|c|c|c|c|c|c|c|c|}
\hline\hline                                     & $B_{unit}$	         & $\beta_e$     	&$\alpha_{MHD}$  & $s$ 		  & $q$ 		& $\kappa$   	& $\delta$  	& $s_{\delta}$   & $s_{\kappa}$\\ [0.5ex] \hline 
77922 (Hyb high-$\delta$)          & 2.86         		& 0.01                & 0.19	    	       & 0.012            & 1.06    	& 1.290   		& 0.028              & 0.028 		& -0.0043          \\
\hline 	
75225 (Hyb low-$\delta$)            & 2.56       		& 0.0097           & 0.26	                 & 0.062           & 1.09    	& 1.357   		& 0.03   	    	& 0.027		& -0.012	         \\
\hline
76679 (BL high-$\delta$)             & 3.76        		& 0.0085           & 0.11      		       & 0.14              & 1.17   	& 1.39   		& 0.041              & 0.045		& 0.014          	\\  
\hline
78682 (BL low-$\delta$)              & 2.90        		& 0.012             & 0.07                        &0.12              & 1.05  		&1.4    		& 0.033              & 0.036		& 0.008	          \\  
\hline
\end{tabular}
\label{table1}
\end{table}

\begin{table}[ht]
\caption{Input parameters for plasma shape and magnetic geometry at $r/a=0.6$.}
\centering
\begin{tabular}{|c|c|c|c|c|c|c|c|c|c|c|}
\hline\hline &		                    $B_{unit}$       & $\beta_e$     &$\alpha_{MHD}$      & $s$ 	    & $q$ 		& $\kappa$   	& $\delta$  	& $s_{\delta}$   & $s_{\kappa}$\\ [0.5ex] \hline 
77922 (Hyb high-$\delta$)          & 3.05       & 0.0049           & 0.16    			& 1.05           & 1.34    	& 1.3   		& 0.063   	    	& 0.1			& 0.055	         \\ 
\hline 	
75225 (Hyb low-$\delta$)           & 2.78        & 0.0044            & 0.21    			& 1.08           & 1.46    	& 1.35   		& 0.065   	    	&0.1			&0.045		\\  
\hline
76679 (BL high-$\delta$)            & 4.06        & 0.0039            & 0.14    			& 0.7             & 1.48   		& 1.43 		& 0.1   	    	&0.15		&0.078               	\\ 
\hline
78682 (BL low-$\delta$)              & 3.10        & 0.006              & 0.2    			&0.62             & 1.3   		& 1.43 		& 0.08   	    	&0.11		&0.052		\\ 
\hline
\end{tabular}
\label{table2}
\end{table}

\begin{table}[ht]
\caption{Input parameters for densities, temperatures and their gradients at $r/a=0.3$.}  
\centering
\begin{tabular}{|c|c|c|c|c|c|c|c|c|c|c|c|c|c|}
\hline\hline
            			       	     & $Z_{eff}$ &  $n_e [10^{19}/m^3]$ &$T_e[keV]$ &$a/L_{ne}$ & $a/L_{Ti}$ & $a/L_{Te}$ &$T_i/T_e$ & $\nu_{ei} (a/c_s)$& $k_{\theta}\rho_{s}$\\ [0.5ex] \hline
77922 (Hyb high-$\delta$)    & 1.53        	  & 5.4 		     	        & 3.9               & 0.66           & 1.34            & 1.38              & 1.48           &        0.028  	    & 0.3               \\ 
\hline
75225 (Hyb low-$\delta$)      & 1.37           & 3.96		     	        & 4.1               & 0.8              & 1.86            & 1.3            	   & 1.77            &        0.02		    & 0.15                  \\ 
\hline         
76679 (BL high-$\delta$)       & 1.83          & 11.0 	    	    	        & 2.9               & 0.22            & 0.84            & 1.43              & 1.14            &        0.11		    & 1.5                  \\ 
\hline
78682 (BL low-$\delta$)        & 1.9           & 6.25	    	    	     	        & 4.1              & -0.18            & 1.13            & 1.                  & 1.05            &        0.03		    & 0.45                  \\ 
\hline
\end{tabular}
\label{table3}
\end{table}

\begin{table}[ht]
\caption{Input parameters for densities, temperatures and their gradients at $r/a=0.6$.}  
\centering
\begin{tabular}{|c|c|c|c|c|c|c|c|c|c|c|c|c|c|}
\hline\hline
            			           & $Z_{eff}$ & $n_e [10^{19}/m^3]$ &$T_e[keV]$ &$a/L_{ne}$   & $a/L_{Ti}$ & $a/L_{Te}$ &$T_i/T_e$ & $\nu_{ei} (a/c_s)$& $k_{\theta}\rho_{s}$\\ [0.5ex] \hline
77922 (Hyb high-$\delta$) & 1.53        & 4.35 		     	    & 2.6                  & 0.94           & 2.1              & 1.16              & 1.22           &   0.051  	    & 0.3               \\ 
\hline
75225 (Hyb low-$\delta$) & 1.37        & 3.96		     	     & 4.1                  & 1.25          & 2.1              & 1.2            	& 1.4            &   0.033	    	    & 0.3                  \\ 
\hline         
76679 (BL high-$\delta$)  & 1.83        & 9.5 	    	    	     & 1.7                  & 0.4             & 2.5              & 1.7                & 1.1            &   0.25	              & 0.3                  \\ 
\hline
78682 (BL low-$\delta$)   & 1.9          & 5.7 	    	    	     & 2.5                  & 1.15           & 2.1              & 2.26              & 1.             &   0.07	              & 0.3                  \\ 
\hline
\end{tabular}
\label{table4}
\end{table}

\clearpage 
\section{Linear instability analysis}
\label{sec:1}
In this section we present the results of our linear stability
analysis with the \gyro code, including both shear- and compressional
magnetic perturbations ($\delta B_\perp\;(=\nabla\times\delta
A_{\parallel})$ and $\delta B_{\parallel}$, respectively).
Drift-kinetic electrons are assumed, and the collisions are modeled
using an energy dependent Lorentz operator.  Both electron-ion and
electron-electron collisions are included in the electron collision
frequency $\nu_e(v)$, and collisions between all ion species are
accounted for. To take \red{the plasma shape into account} we have
used a Miller-type local equilibrium model available in \gyro, see
Refs. \cite{waltz09,candy}\red{.}

Typical resolution parameters used in our linear analysis are as
follows: 40 radial grid points, 12 parallel orbit mesh points ($\times
2$ signs of parallel velocity), 16 pitch angles, and 8 energies.

\begin{figure}[htbp]
\begin{center}
 \includegraphics[width=0.8\textwidth]{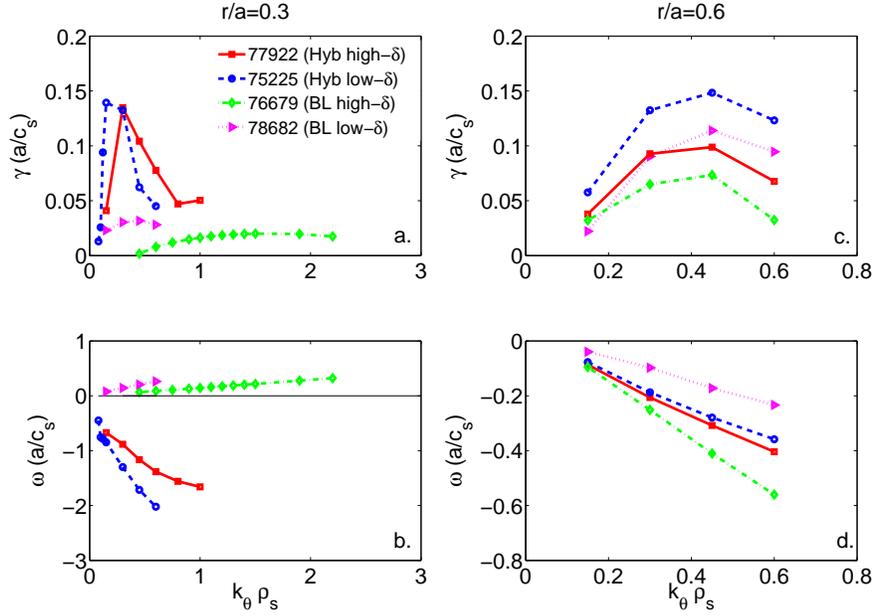}
\caption{Imaginary and real frequency of the most linearly unstable
  modes as functions of $k_{\theta}\rho_s$ (a,b) at $r/a=0.3$, and
  (c,d) at $r/a=0.6$. Solid (red) lines correspond to the values for
  77922, dashed (blue) lines for 75225, dashed-dotted (green) lines
  for 76679, and dotted (mauve) lines for 78682 discharges.}
\label{fig4}
\end{center}
\end{figure}

\begin{figure}[htbp]
\begin{center}
 \includegraphics[width=0.83\textwidth]{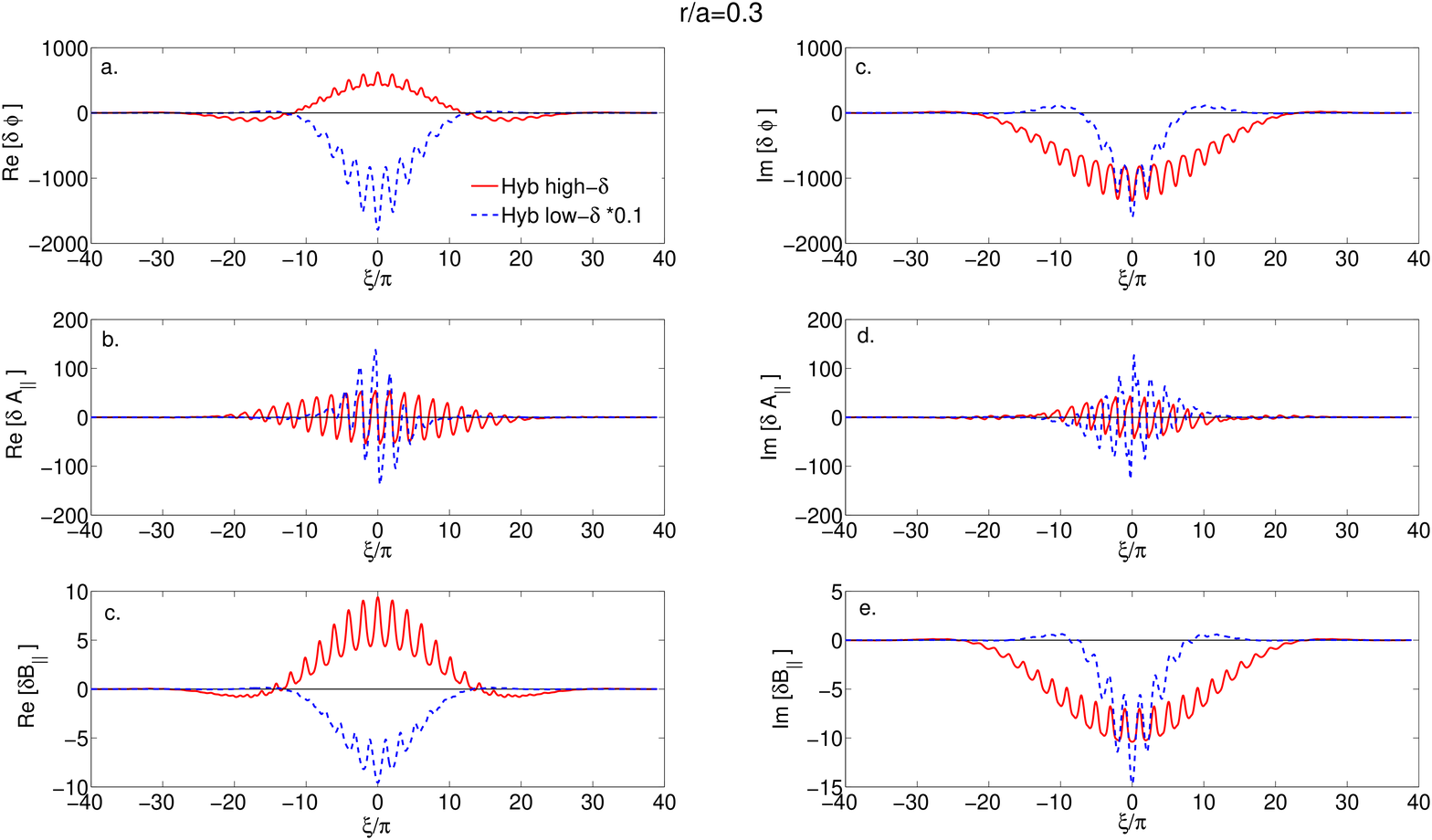}\\
 \includegraphics[width=0.83\textwidth]{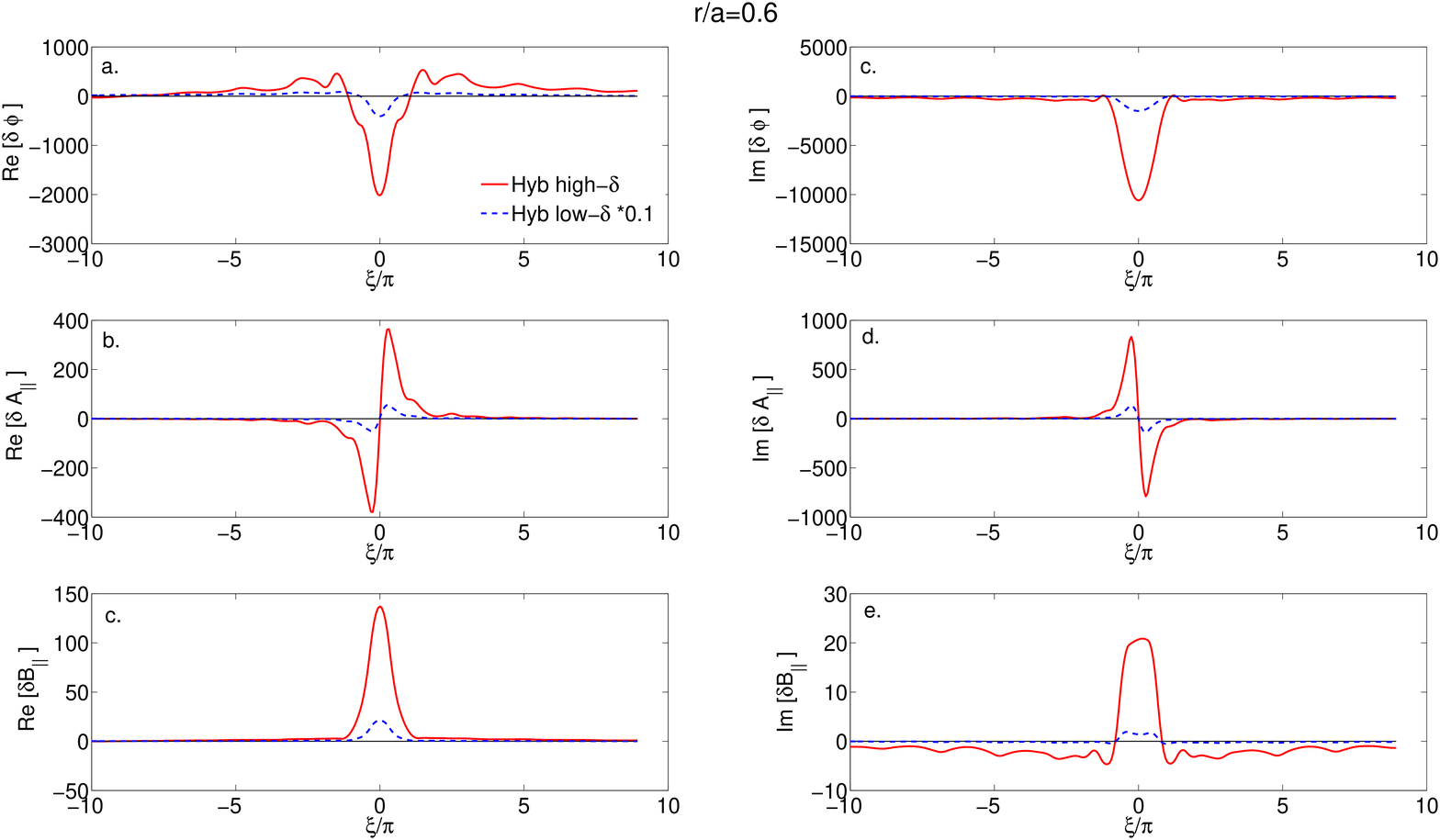}
\caption{Linear parallel mode structures of $\delta \phi$, $\delta
  A_{\parallel}$, and $\delta B_{\parallel}$ as functions of the
  normalized extended poloidal angle $\xi/\pi$ for hybrid plasmas. Top
  figures show the values for $r/a=0.3$, and bottom figures show the
  values for $r/a=0.6$. Solid (red) lines correspond to the values for
  77922, dashed (blue) lines for 75225.}
\label{fig5}
\end{center}
\end{figure}

\begin{figure}[htbp]
\begin{center}
 \includegraphics[width=0.83\textwidth]{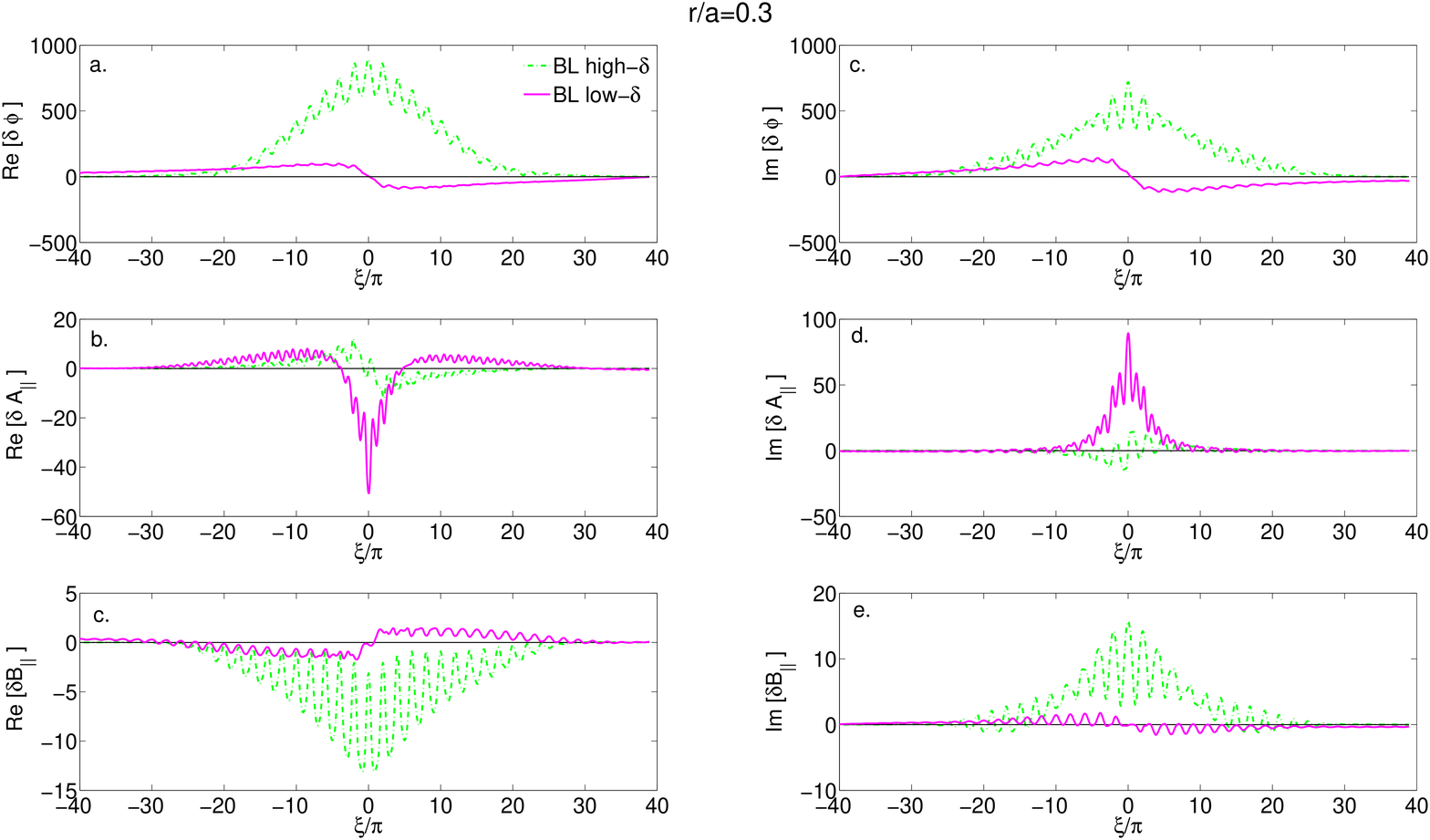}\\
 \includegraphics[width=0.83\textwidth]{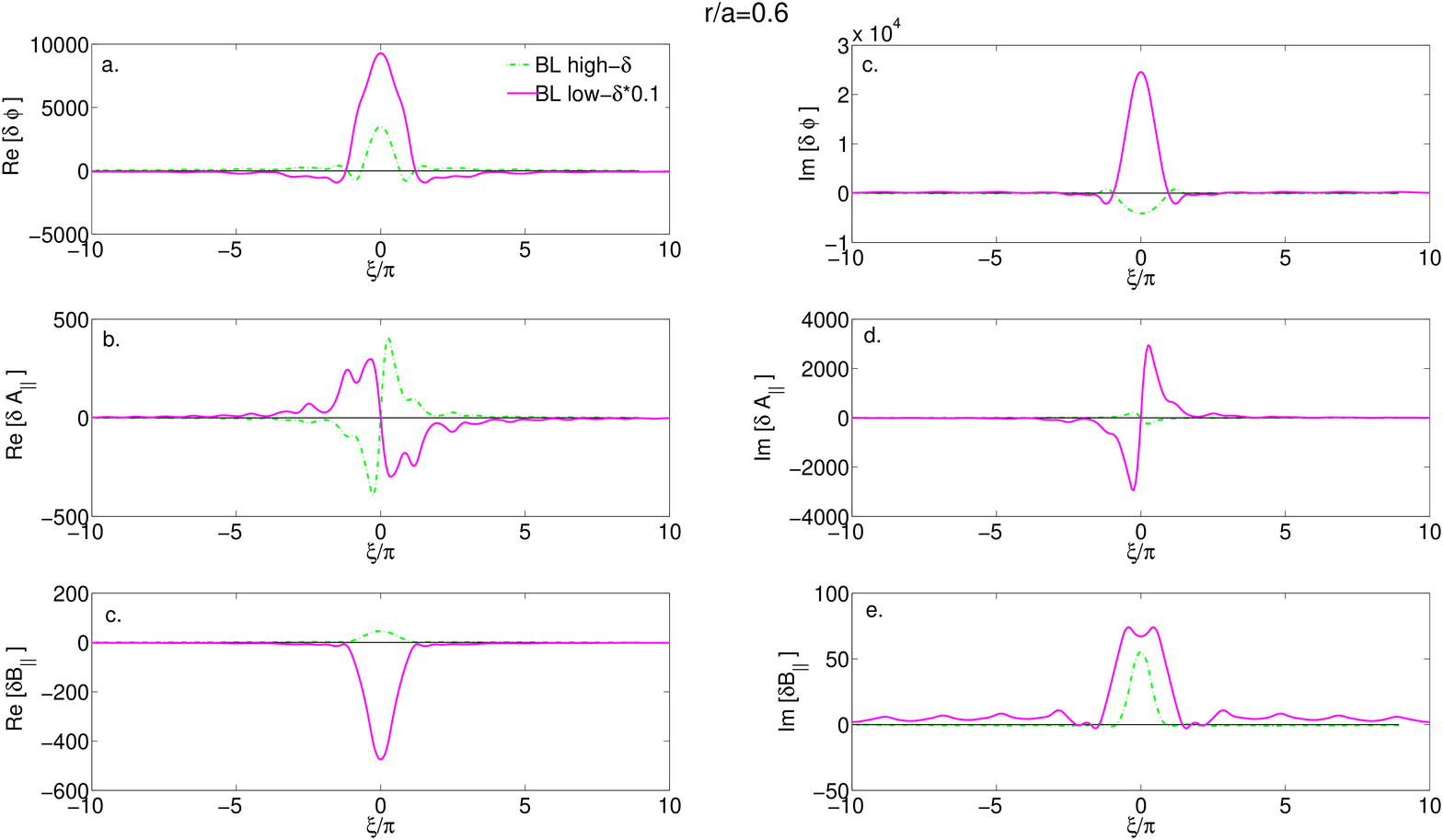}
\caption{Linear parallel mode structures of $\delta \phi$, $\delta
  A_{\parallel}$, and $\delta B_{\parallel}$ as functions of the
  normalized extended poloidal angle $\xi/\pi$ in baseline
  plasmas. Top figures show the values for $r/a=0.3$, and bottom
  figures show the values for $r/a=0.6$. Dashed-dotted (green) lines
  for 76679, and dotted (mauve) lines for 78682 discharges.}
\label{fig5-1}
\end{center}
\end{figure}

Figure \ref{fig4} shows the growth rates \red{$\gamma$} and real
frequencies \red{$\omega$} of the most unstable linear modes as
functions of $k_{\theta}\rho_s$ for the selected discharges at both
radii \red{(the values are given in $c_s/a$ units, with
  $c_s=(T_e/m_i)^{1/2}$ the ion sound speed)}. As shown in
Fig. \ref{fig4}(a) and (b), for the inner core region at $r/a=0.3$, there
is a difference in the nature of the underlying unstable modes between
the hybrid and baseline plasmas. The most unstable modes in both
hybrid plasmas are kinetic ballooning modes (KBMs), propagating in the
ion diamagnetic direction (negative $\omega$), while in the baseline
plasmas we find unstable \red{trapped electron mode (TEM) and
  micro-tearing mode (MTM)} in the high-$\delta$ and the low-$\delta$
cases, respectively; these modes propagate in the electron diamagnetic
direction (positive $\omega$). However, in the outer core, at
$r/a=0.6$, \red{ion temperature gradient (ITG) modes} are the most
unstable \red{instabilities} in all selected discharges, see
Fig. \ref{fig4}(c) and (d). The $\delta \phi$, $\delta A_{\parallel}$ and
$\delta B_{\parallel}$ components of the parallel eigenmode structure
are shown in Figs. \ref{fig5} and \ref{fig5-1}. The eigenfunctions are
normalized so that $\delta A_{\parallel}(\xi=0)$ is unity. The MTM
signature is distinguished by the odd (even) parity of the
eigenfunction in $\delta \phi$ ($\delta A_{\parallel}$), and
correspondingly for KBMs the even (odd) parity of the eigenfunction in
$\delta \phi$ ($\delta A_{\parallel}$) eigenmodes
\cite{hatchpop2013}. Due to very low magnetic shear, the eigenmodes
are very elongated along the field line in the inner core radius
  ($r/a=0.3$) and therefore a very high radial resolution was needed
  (we used 40 radial grid pints) to resolve the modes.  The above
  mentioned identification of the modes, is not purely based on the
  parities of their eigenmodes, but also on their dependencies on the
  various plasma gradient scaling lengths which will be discussed in
  the following sections.
 
\begin{figure}[htbp]
\begin{center}
 \includegraphics[width=0.9\textwidth]{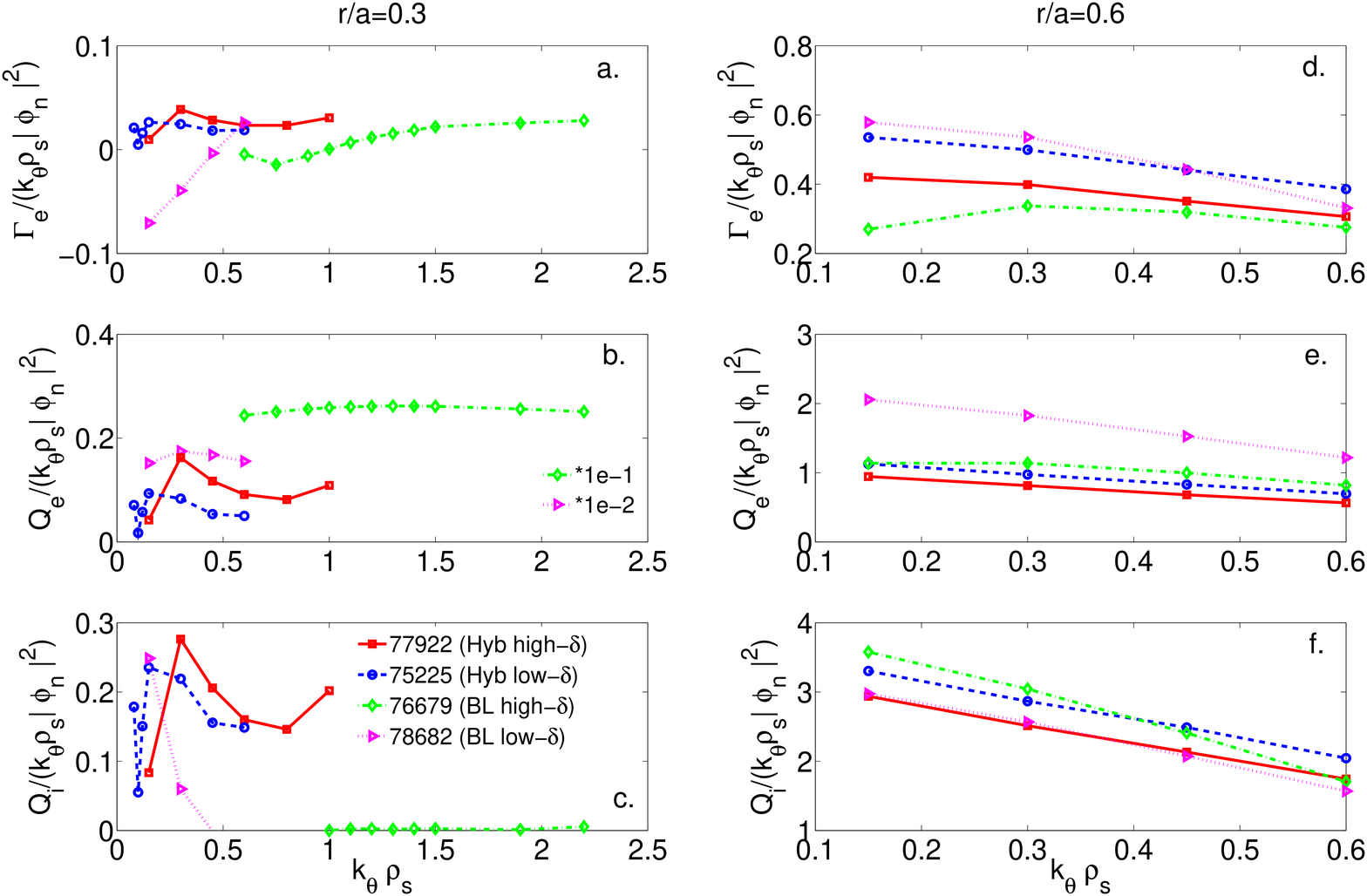}
\caption{Normalized linear particle (a,d), and heat fluxes of
  electrons (b,e) and ions (c,f); left figures at $r/a=0.3$, and right
  figures at $r/a=0.6$. Solid (red) lines correspond to the values for
  77922, dashed (blue) lines for 75225, dashed-dotted (green) lines
  for 76679, and dotted (mauve) lines for 78682 discharges. 
  \newnew{Fluxes are normalised to $k_{\theta} \rho_s |\phi_n|^2$}.}
\label{fig4-1}
\end{center}
\end{figure}

Figure \ref{fig4-1} \red{illustrates} the normalized linear heat and
particle fluxes as functions of $k_{\theta}\rho_{s}$ for all the
selected discharges. For the inner core radius, as seen in
Fig. \ref{fig4-1}(a), KBMs in hybrid plasmas generate positive
(outward) particle flux with the maximum located at the
$k_{\theta}\rho_{s}$ corresponding to the maximum of their growth
rates (see Fig. \ref{fig4}(a)). In the high-$\delta$ baseline TEMs
generate negative (inward) particle flux at low $k_{\theta}\rho_{s}$
but the maximum particle flux is positive and located at
$k_{\theta}\rho_{s}$ corresponding to the maximum growth rate (see
Fig. \ref{fig4}(a)). In the low-$\delta$ baseline however, the MTM
driven particle flux varies from negative to positive as
$k_{\theta}\rho_{s}$ increases, and at the position of the
$k_{\theta}\rho_{s}$ corresponding to the maximum growth rate the flux
is close to zero. This is in agreement with the previous linear
\cite{smoradipop2013} and non-linear
\cite{GuttenfelderPRL2011,GuttenfelderPoP2012NL} works where it was
also shown that particle fluxes driven by MTMs are negligible. 

The main channel of transport for KBMs in hybrid plasmas is the heat
flux with slightly higher levels obtained for the ion heat flux, as
shown in Figs. \ref{fig4-1}(b) and (c). As seen in these figures, the
electron heat flux generated by TEM modes in low-$\delta$ baseline
plasma is significantly higher than those generated by KBMs in the
hybrids whereas the ion heat flux due to TEMs is negligibly small and
close to zero. Note that for more clear view the values of the
electron heat fluxes for the baseline plasmas in Fig. \ref{fig4-1}(b)
have been rescaled where the scaling parameters are $0.1$ and
  $0.01$ for the high and low-$\delta$ baseline plasmas,
respectively. In the low-$\delta$ baseline plasma, the MTMs generate
the highest level of the electron heat flux compared to the other
three plasmas, as seen in Fig. \ref{fig4-1}(b), while the ion heat flux
as shown in Fig. \ref{fig4-1}(c), is negligibly small. 

However, note that such high values of the MTM-driven electron heat flux is
  partly an artifact of the normalization of the linear fluxes, since
  the ratio of the perturbed electrostatic potential $\phi$ and the
  parallel vector potential $A_\|$ is smaller for MTMs than for the
  other modes, while the electron heat flux is mostly $A_\|$-driven
  ``flutter'' transport in that case. 

For the outer core radius, as seen in Figs. \ref{fig4-1}(d-f), however,
the particle and heat fluxes are driven by ITG modes in all the
selected plasmas and show similar trends as $k_{\theta}\rho_{s}$
varies, with similar values obtained for all four studied
discharges. In all the cases the particle fluxes generated by ITGs are
positive and outward directed, with higher levels compare to those for
the inner core radius, see Fig. \ref{fig4-1}(d). Similarly the electron
and ion heat fluxes generated by ITGs are higher than those obtained
at $r/a=0.3$ with the exception of the electron heat flux driven by
MTMs in the low-$\delta$ baseline plasma, which is an order of
magnitude higher.
 
\subsection{Sensitivity scans for $\beta_e$ and $\alpha_{MHD}$}

\begin{figure}[htbp]
\begin{center}
 \includegraphics[width=0.9\textwidth]{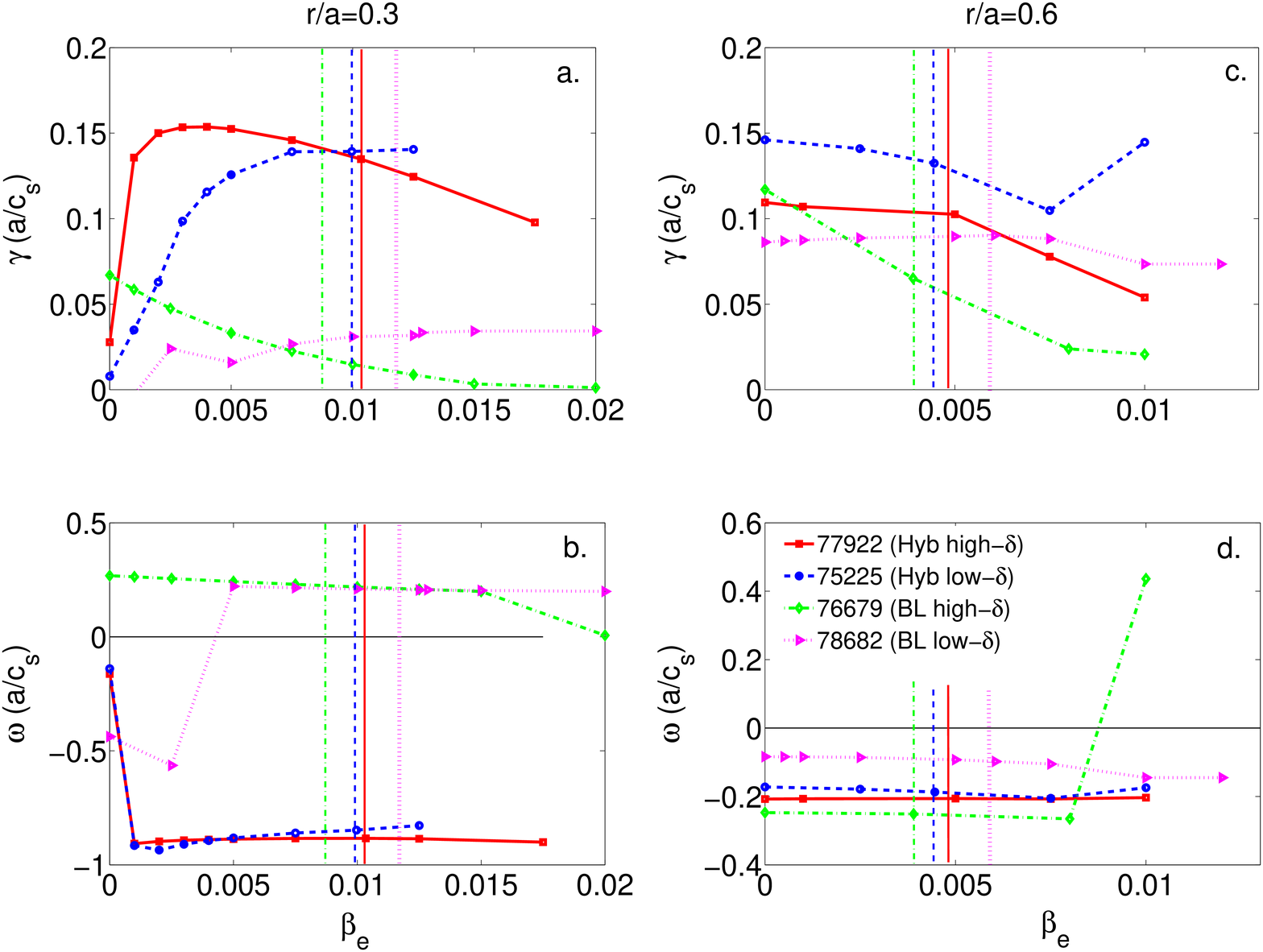} 
\caption{Imaginary and real frequency of the most linearly unstable
  modes as functions of $\beta_e$ with self-consistent variation of
  the $\alpha_{MHD}, $ (a,b) at $r/a=0.3$, and (c,d) at
  $r/a=0.6$. Solid (red) lines correspond to the values for 77922,
  dashed (blue) lines for 75225, dashed-dotted (green) lines for
  76679, and dotted (mauve) lines for 78682 discharges. The straight
  lines represent the experimental values for $\beta_e$ colour coded
  similarly to the corresponding curves.}
\label{fig6}
\end{center}
\end{figure}

\begin{figure}[htbp]
\begin{center}
 \includegraphics[width=0.43\textwidth]{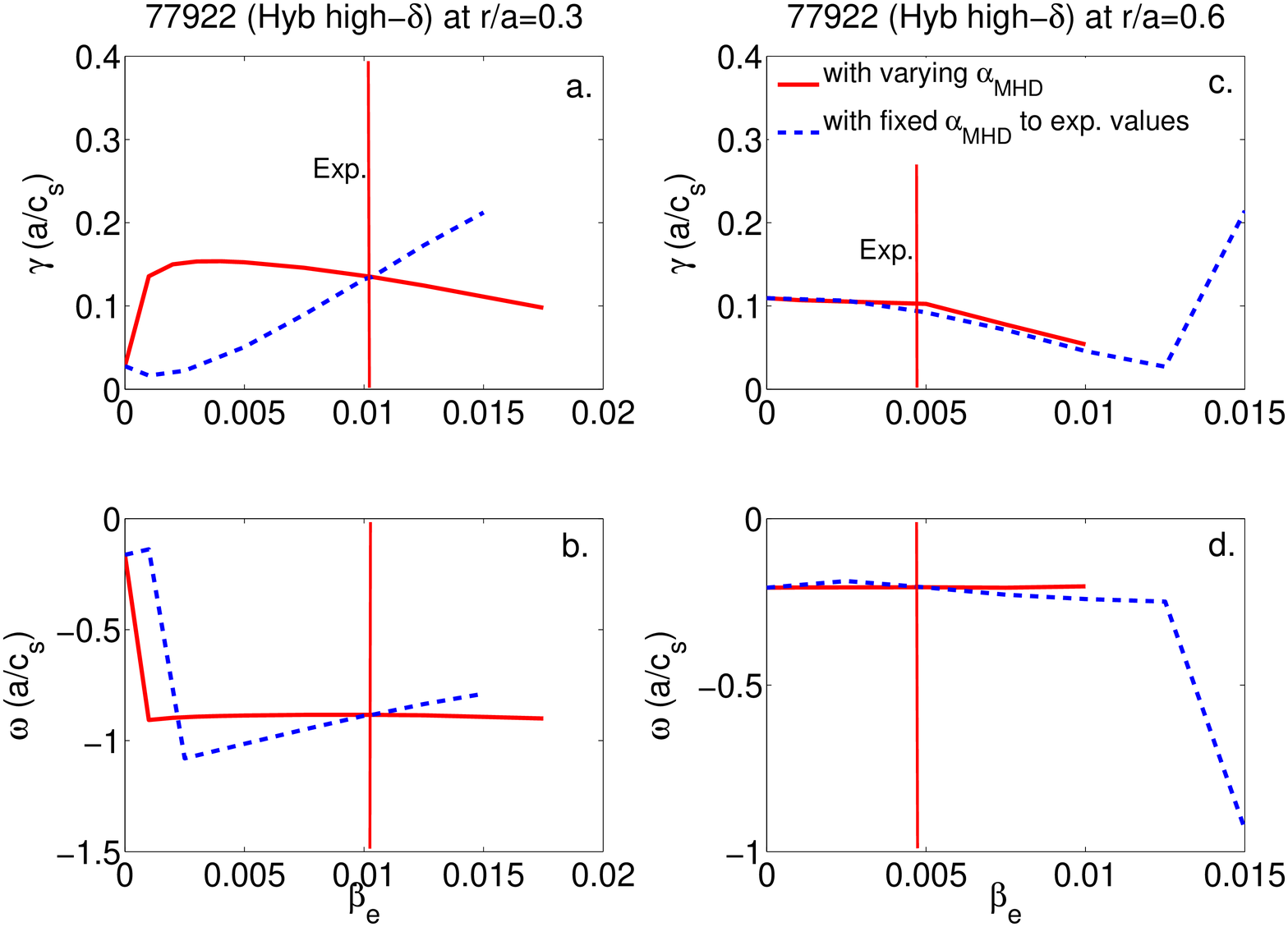} \includegraphics[width=0.43\textwidth]{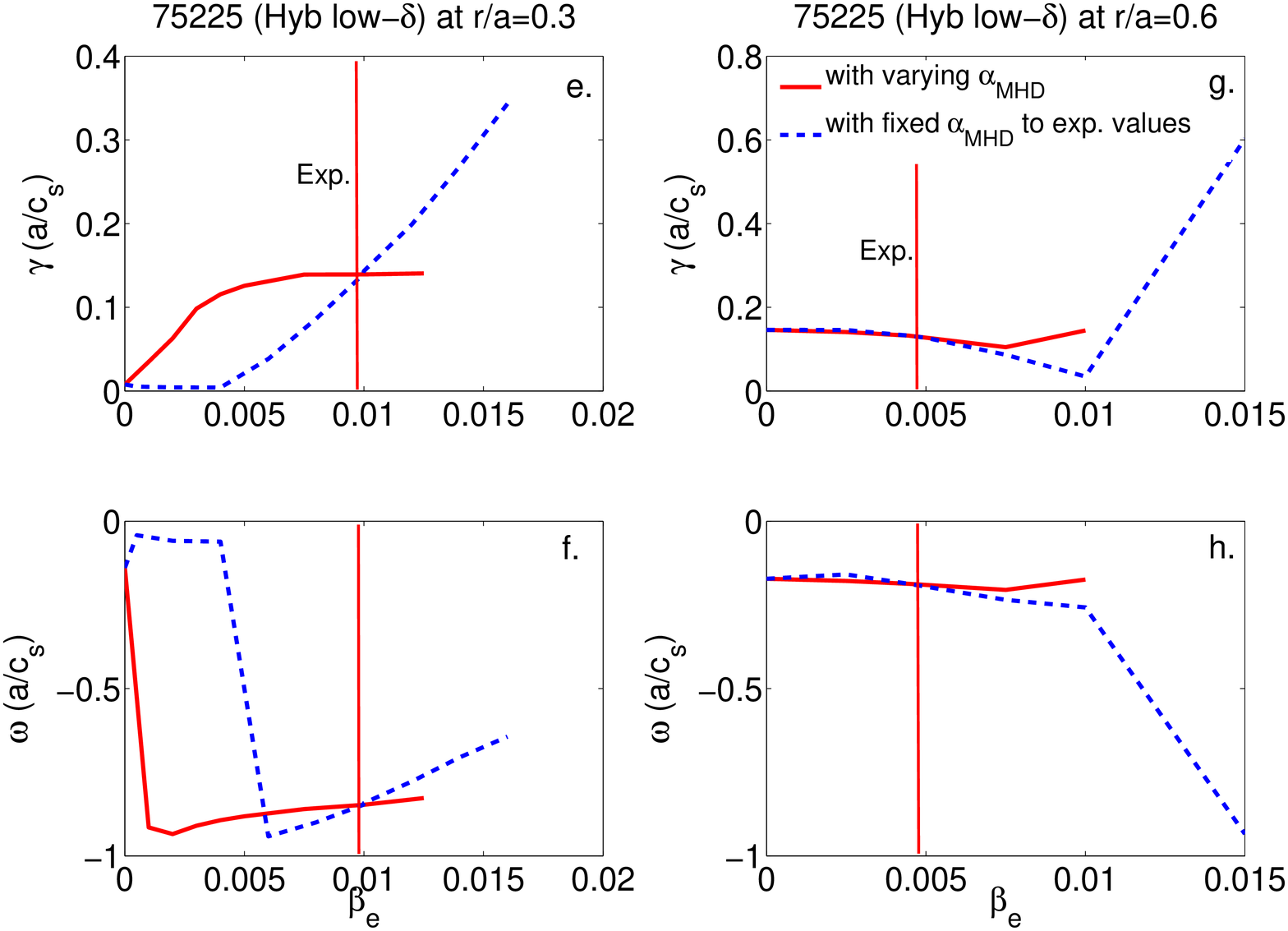}\\
 \includegraphics[width=0.43\textwidth]{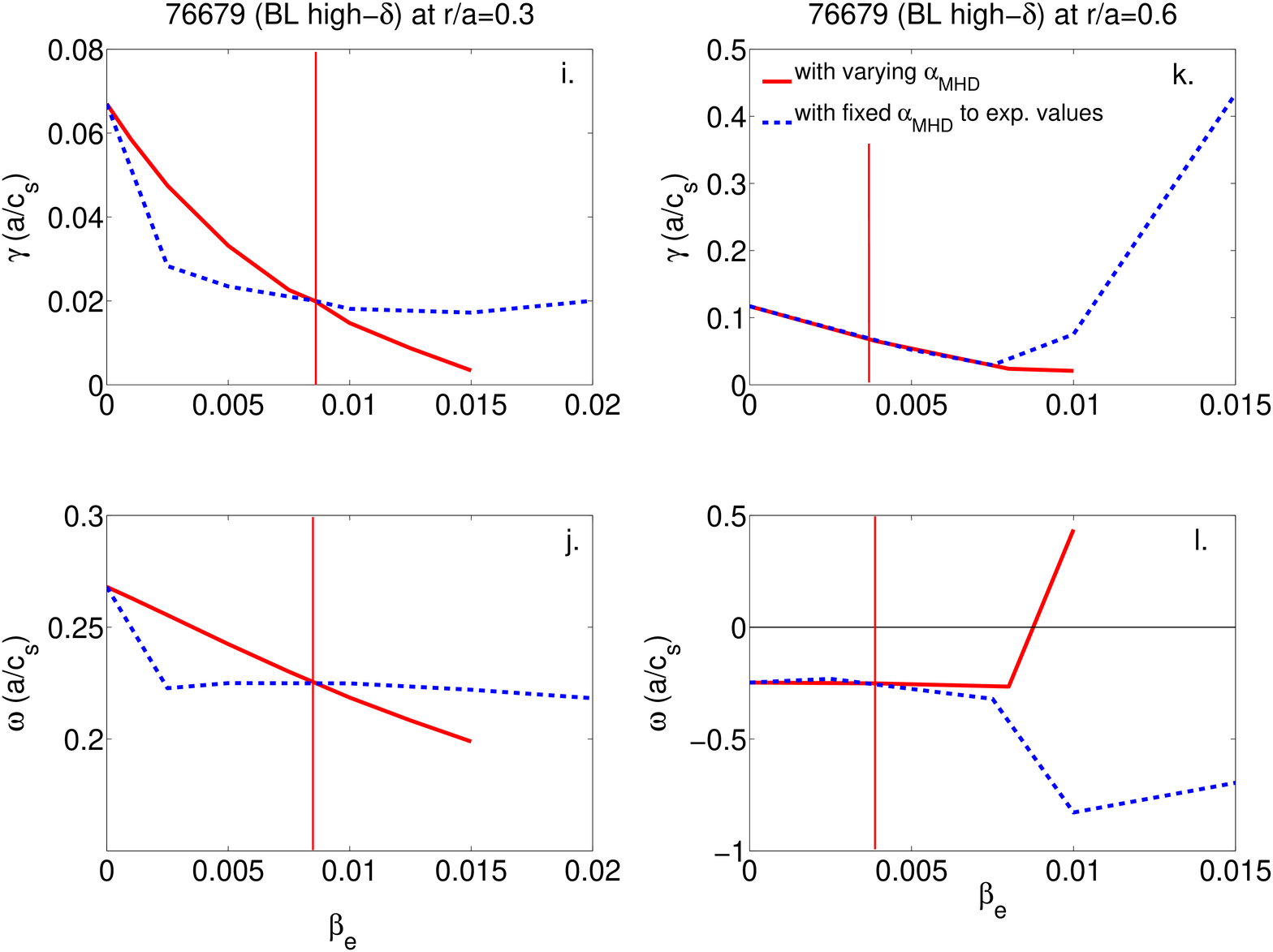} \includegraphics[width=0.43\textwidth]{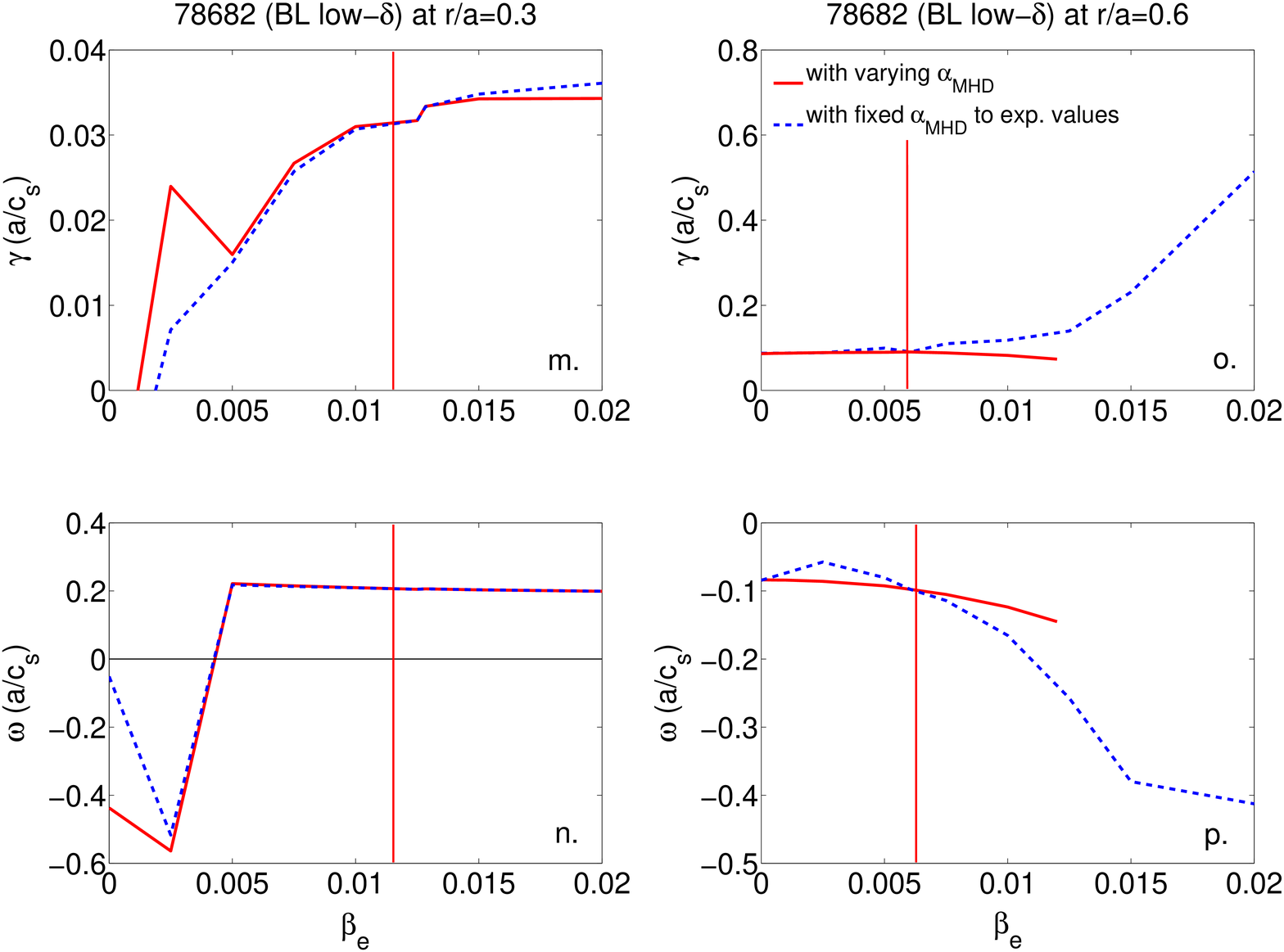}
\caption{Imaginary and real frequency of the most unstable modes as
  functions of $\beta_e$ with self-consistent variation of the
  $\alpha_{MHD}$ solid (red) lines, and with fixed $\alpha_{MHD}$
  dashed (blue) lines. (a,b) at $r/a=0.3$ and (c,d) at $r/a=0.6$ for
  77922 discharge, (e,f) at $r/a=0.3$ and (g,h) at $r/a=0.6$ for 75225
  discharge. (i,j) at $r/a=0.3$ and (k,l) at $r/a=0.6$ for 76679
  discharge. (m,n) at $r/a=0.3$ and (o,p) at $r/a=0.6$ for 78682
  discharge.}
\label{fig6-1}
\end{center}
\end{figure}

In order to analyze the importance of the plasma pressure on the
instability of electromagnetic modes, we performed $\beta_e$ scans for
the selected discharges at the two core radii, see Fig. \ref{fig6},
where the experimental values of $\beta_e$ are shown by vertical lines
with similar colour coding to their corresponding curves. As seen in
Figs. \ref{fig6}(a) and (b) for the inner core radius in the hybrid
plasmas (high-$\delta$: red solid line, and low-$\delta$: blue dashed
line), in the absence of the electromagnetic effects ($\beta_e=0$), we
find that the most unstable modes are ITGs, and by including the
electromagnetic effects the ITGs are stabilized, while the KBMs are
destabilized and become the dominant instability with a very low
$\beta_e$ threshold: $\beta_e < 0.1\%$. In both hybrid plasmas, the
KBM growth rate shows a non-monotonic change as the $\beta_e$ is
increased. After an initial phase of destabilization by $\beta_e$, the
KBMs become stabilized by further increase in $\beta_e$. The
stabilization is due to the increase in the $\alpha_{MHD}$ with
$\beta_e$ (see Eq.~(\ref{alpha})), and is demonstrated in
Fig. \ref{fig6-1}, where the impact of $\alpha_{MHD}$ on the
stabilization of the modes is examined. In this figure red (solid)
lines represent the scans with self-consistent variation of the
$\alpha_{MHD}$ with $\beta_e$ and the blue (dashed) lines represent
the scans in which by varying the $c_{p}$ parameter as $\beta_e$
varies (see Eq.~(\ref{alpha})) we ensure that $\alpha_{MHD}$ is kept
constant and equal to the corresponding experimental value given in
tables \ref{table1} and \ref{table2}. As seen in Figs. \ref{fig6-1}(a,b)
(high-$\delta$ hybrid) and (e,f) (low-$\delta$ hybrid) at $r/a=0.3$ and
bellow the experimental $\beta_e$ values marked by the straight lines,
the onset of the mode is moved to higher $\beta_e$, since the
$\alpha_{MHD}$ is high and therefore the modes are stabilized. As
$\beta_e$ increases beyond the experimental values and because
$\alpha_{MHD}$ is kept constant the growth rates increase strongly
without any further stabilization.

As shown in Figs. \ref{fig6}(a) and (b), in the high-$\delta$ baseline
plasma (green dashed dotted lines) the growth rate of the TEM mode
decreases with increasing $\beta_e$, but the TEM remains the dominant
instability throughout the $\beta_e$ scan. The stabilization due to
$\alpha_{MHD}$ for the TEM mode is strong as seen in
Figs. \ref{fig6-1}(i) and (j). The MTMs in the low-$\delta$ baseline
plasma (mauve dotted lines) shown in Figs. \ref{fig6}(a) and (b), are
destabilized for $\beta_e>0.5\%$ and their growth rate increases with
$\beta_e$, however above the experimental level of $\beta_e$ the mode
growth rate does not increase significantly. \red{There} is very little
stabilization due to $\alpha_{MHD}$ on the MTM modes, as seen in
Figs. \ref{fig6-1}(m) and (n).
 
At the outer core radius ($r/a=0.6$) ITG is the most unstable mode in
all the selected plasmas, which is slightly stabilized with increasing
$\beta_e$. The ITG growth rates are comparable to the values obtained
in the absence of the electromagnetic effects ($\beta_e=0$), see
Figs. \ref{fig6}(c) and (d). Also, as seen in Figs. \ref{fig6-1}(c,d)
(high-$\delta$ hybrid), g,h (low-$\delta$ hybrid), k,l (high-$\delta$
baseline) and (o,p) (low-$\delta$ baseline), there is very little
stabilization from $\alpha_{MHD}$ on the ITG modes and in the presence
of constant $\alpha_{MHD}$, while the KBMs are destabilized as
$\beta_e$ is increased above the experimental values.

\subsection{Sensitivity to profile gradients and collisionality}
\begin{figure}[htbp]
\begin{center}
 \includegraphics[width=0.45\textwidth]{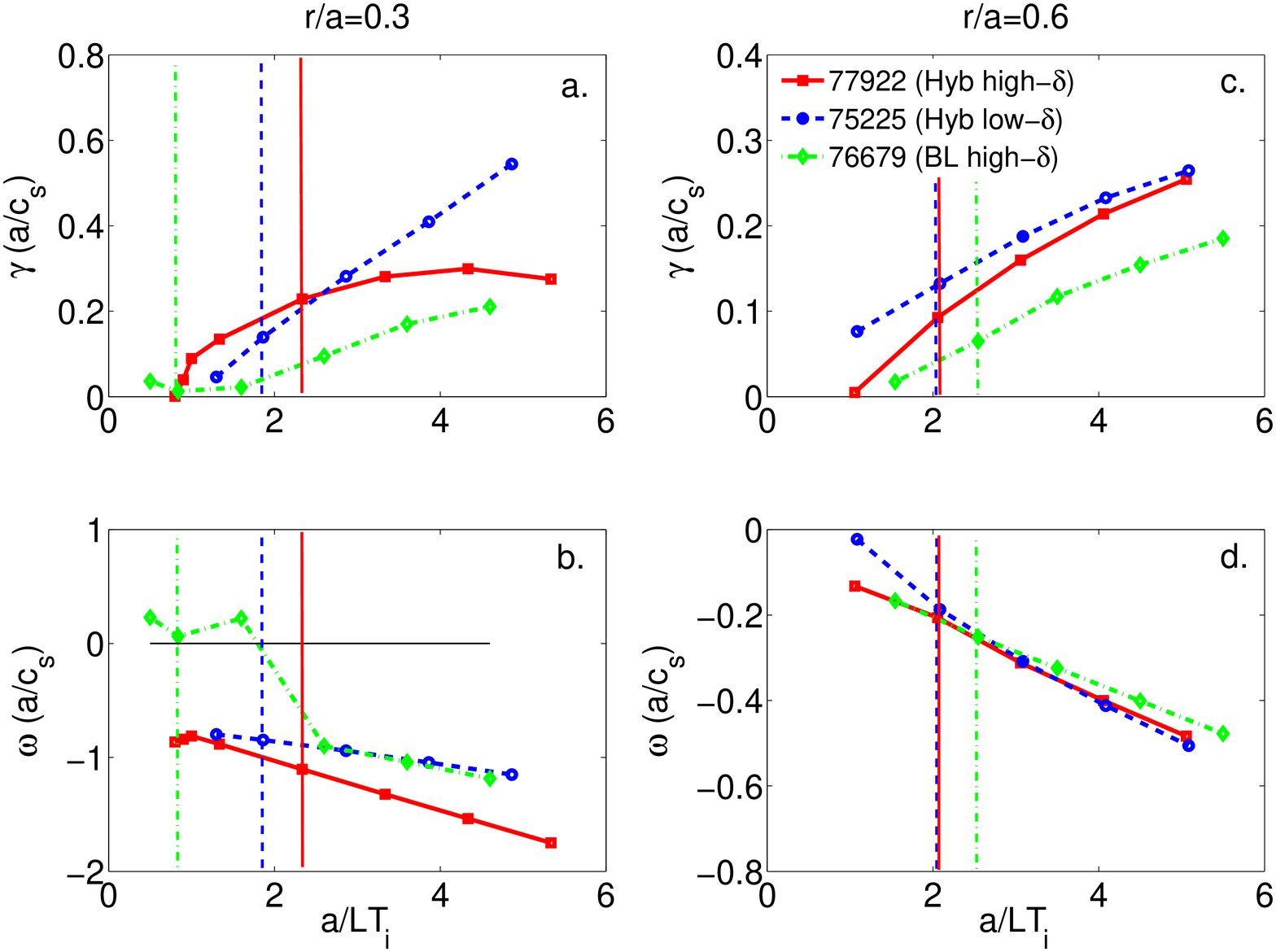} \includegraphics[width=0.45\textwidth]{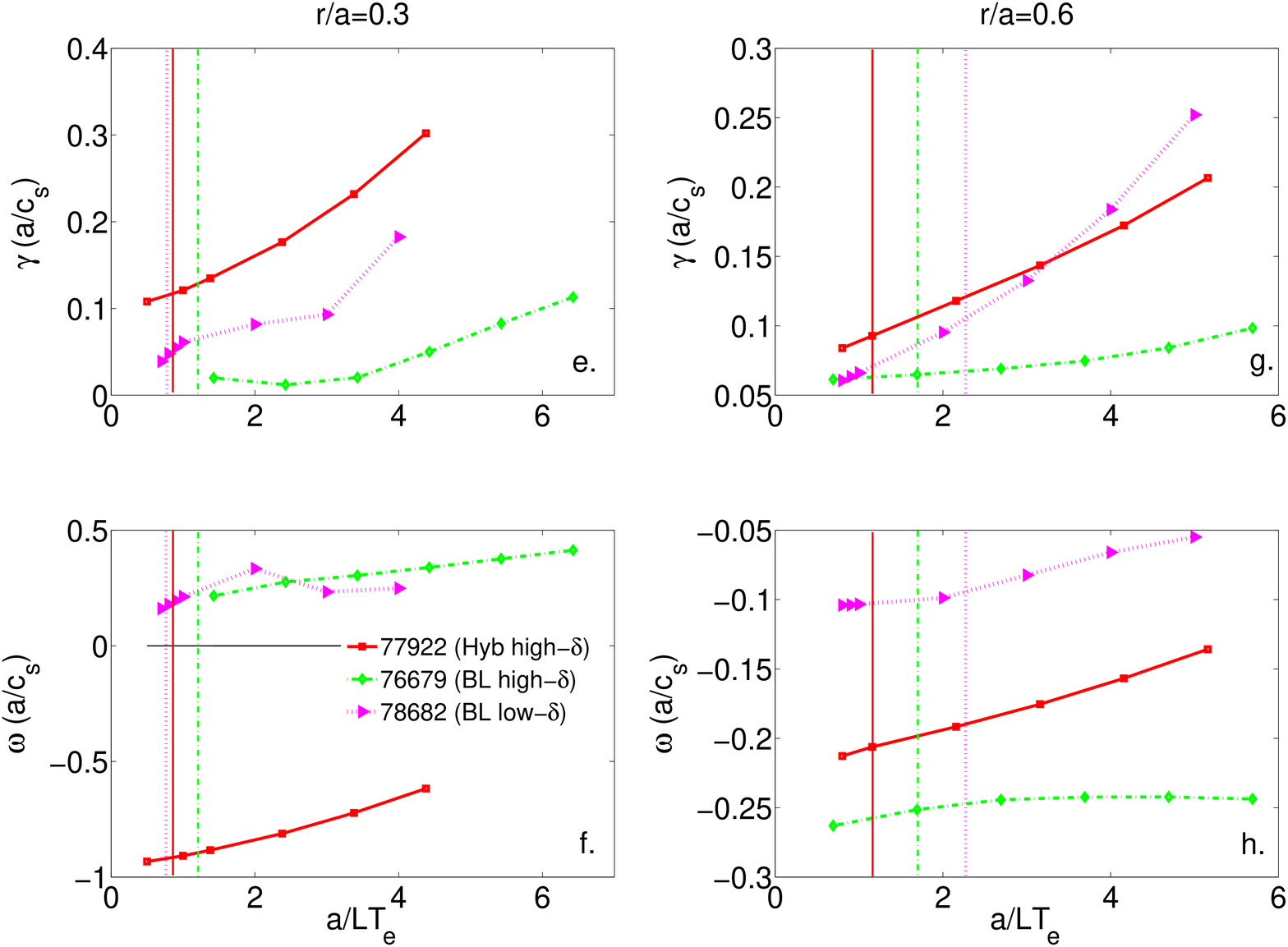}
\caption{Imaginary and real frequency of the most linearly unstable
  modes as functions of $a/L_{Ti}$ (left), and $a/L_{Te}$
  (right). (a,b) and (e,f) at $r/a=0.3$, (c,d) and (g,h) at
  $r/a=0.6$. Solid (red) lines correspond to the values for 77922,
  dashed (blue) lines for 75225, dashed-dotted (green) lines for
  76679, and dotted (mauve) lines for 78682 discharges. The straight
  lines represent the experimental values for $a/L_{Ti,e}$ colour
  coded similarly to the corresponding curves.}
\label{fig7}
\end{center}
\end{figure}

In this section the results of our investigation on the impact of
temperature and density gradients, and collisionality on the
stability of the modes in the selected discharges are
presented. Figures \ref{fig7}(a-h) show the $a/L_{Ti,e}$ sensitivity
studies at two radii. In all these scans $\alpha_{MHD}$ is kept
constant and equal to the corresponding experimental values given in
tables \ref{table1} and \ref{table2}. As seen in these figures, by
increasing $a/L_{Ti}$ the KBM (at $r/a=0.3$) and ITG (at $r/a=0.6$)
are further destabilized, because $a/L_{Ti}$ is a drive for these 
modes. However, within 20 to 30\% variation of $a/L_{Ti}$ the most
unstable modes remain unchanged.

On the other hand, for the TEM mode in high-$\delta$ baseline plasma
(green dashed-dotted line), shown in Figs. \ref{fig7}(a) and (b), for
$a/L_{Ti}$ similar to the experimental values in hybrid plasmas (shown
with red and blue straight lines) the most unstable mode switches from
a TEM to a KBM mode when the ion temperature gradient is
  increased sufficiently. This change is more visible in
\red{Fig.} \ref{fig7}(b) were the real frequency of the mode changes from
positive (electron direction) to negative (ion direction) with
comparable values as in the hybrid KBMs (red solid and blue dashed
lines). For the MTMs in low-$\delta$ baseline plasma (mauve dotted
line), shown in Figs. \ref{fig7}(e) and (f), increasing $a/L_{Te}$ results
in destabilization of MTMs. However going to $a/L_{Te}$ highly above
the experimental level (represented by the mauve straight line)
switches the most unstable mode from a MTM to a TEM, more clearly seen
in Fig. \ref{fig7}(f) were the real frequency shows a discontinuity with
a significant reduction.

\begin{figure}[htbp]
\begin{center}
\includegraphics[width=0.8\textwidth]{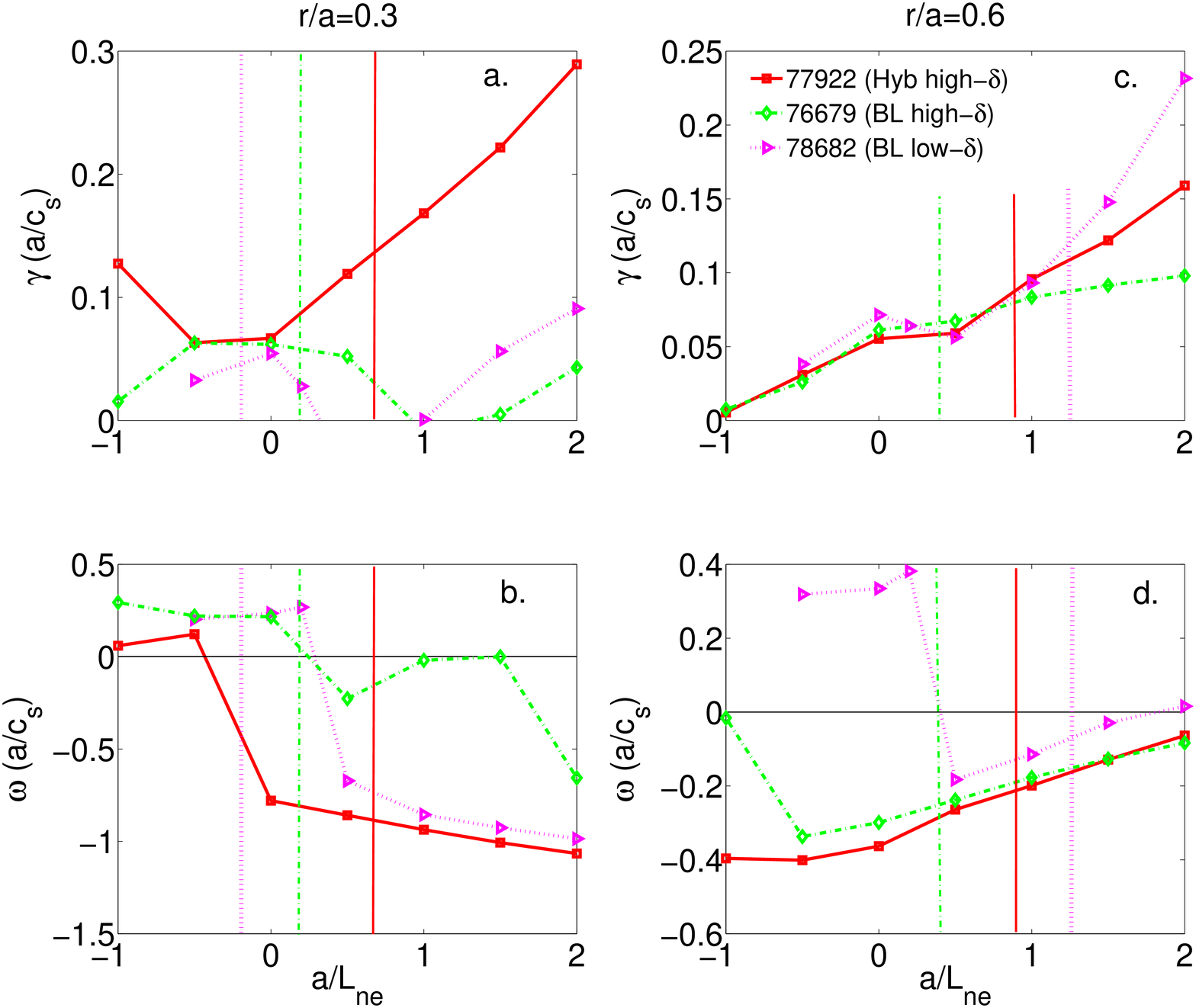}
\caption{Imaginary and real frequency of the most linearly unstable
  modes as functions of $a/L_{ne}$ (right). (a,b) at $r/a=0.3$, (c,d)
  at $r/a=0.6$. Solid (red) lines correspond to the values for 77922,
  dashed-dotted (green) lines for 76679, and dotted (mauve) lines for
  78682 discharges. The straight lines represent the experimental
  values for $a/L_{ne}$ colour coded similarly to the corresponding
  curves.}
\label{fig7_1}
\end{center}
\end{figure}

The sensitivity scans for density gradient scaling lengths,
$a/L_{ne}$, are shown in Figs. \ref{fig7_1}(a-d). Here, quasi-neutrality
is enforced by varying $a/L_{Ni}$ as $a/L_{ne}$ varies, while keeping
the impurity density gradients fixed. In all these scans
$\alpha_{MHD}$ is kept constant and equal to the corresponding
experimental values given in tables \ref{table1} and \ref{table2}. For
the high-$\delta$ hybrid plasma studied here (red solid lines in
Figs. \ref{fig7_1}(a) and (b), and at $r/a=0.3$, a mode change from TEM
to KBM is observed when varying the $a/L_{ne}$ from negative to
positive values (corresponding to a change in density profile from
hallow to peaked, respectively). The mode change is observed around
$a/L_{ne}=0$ (corresponding to a flat density profile), and the KBM
growth rate is increased as $a/L_{ne}$ is increased i.e. the density
profile becomes more peaked. Thus, in the inner core region of the
hybrid plasma studied here, KBMs are expected to remain the dominant
instability for a flat to peaked density profile. However, if the
profile becomes hallow, the most unstable mode changes significantly
and therefore, the expected transport characteristics are very
different as well. No mode change is observed for the outer core
radius, shown in Figs. \ref{fig7_1}(c) and (d), and the main unstable
mode remains an ITG mode for the whole $a/L_{ne}$ range considered
with increasing growth rate as $a/L_{ne}$ is increased. Therefore, in
this region further peaking of the density profile can result in an
increase of the ITG driven transport.

In the high-$\delta$ baseline plasma (green dashed lines in
Figs. \ref{fig7_1}(a) and (b) the characteristics of the most unstable
mode depends strongly on the density gradient scaling length. For the
$a/L_{ne}$ bellow the experimental values (shown by the straight green
dashed lines) TEMs remain the dominant instability however, as the
$a/L_{ne}$ is increased i.e. more peaked density profile, the mode
changes from a TEM to ITG and then to KBM. For the outer core region,
similarly to the hybrid plasma case, the most dominant mode remains an
ITG mode for the whole range of the scan with the growth rate slightly
increasing as $a/L_{ne}$ is increased.

In the low-$\delta$ baseline plasma (mauve dashed-dotted lines in
Figs. \ref{fig7_1}(a) and (b) the MTMs remains the dominant instability
for a range of $-0.2<a/L_{ne}<0.2$, which correspond to a change from
a slightly hallow to flat to slightly peaked density profile. Clearly
$a/L_{ne}$ is not a strong and necessary drive for the MTMs, as there
is a finite MTM growth rate at $a/L_{ne}=0$. However, as $a/L_{ne}$ is
further increased the mode changes from an MTM to a KBM. A transition
from MTM to KBM with increasing $a/L_{ne}$ has also been reported
previously in
Refs. \cite{smoradipop2013,GuttenfelderPoP2012L,DickinsonPRL}. At the
outer core radius, for both baseline plasmas, a mode change from TEM
to ITG is observed as the $a/L_{ne}$ varies from negative to positive
with a stronger variation seen in the case of the low-$\delta$
baseline plasma, see Figs. \ref{fig7_1}(c) and (d).

\begin{figure}[htbp]
\begin{center}
 \includegraphics[width=0.8\textwidth]{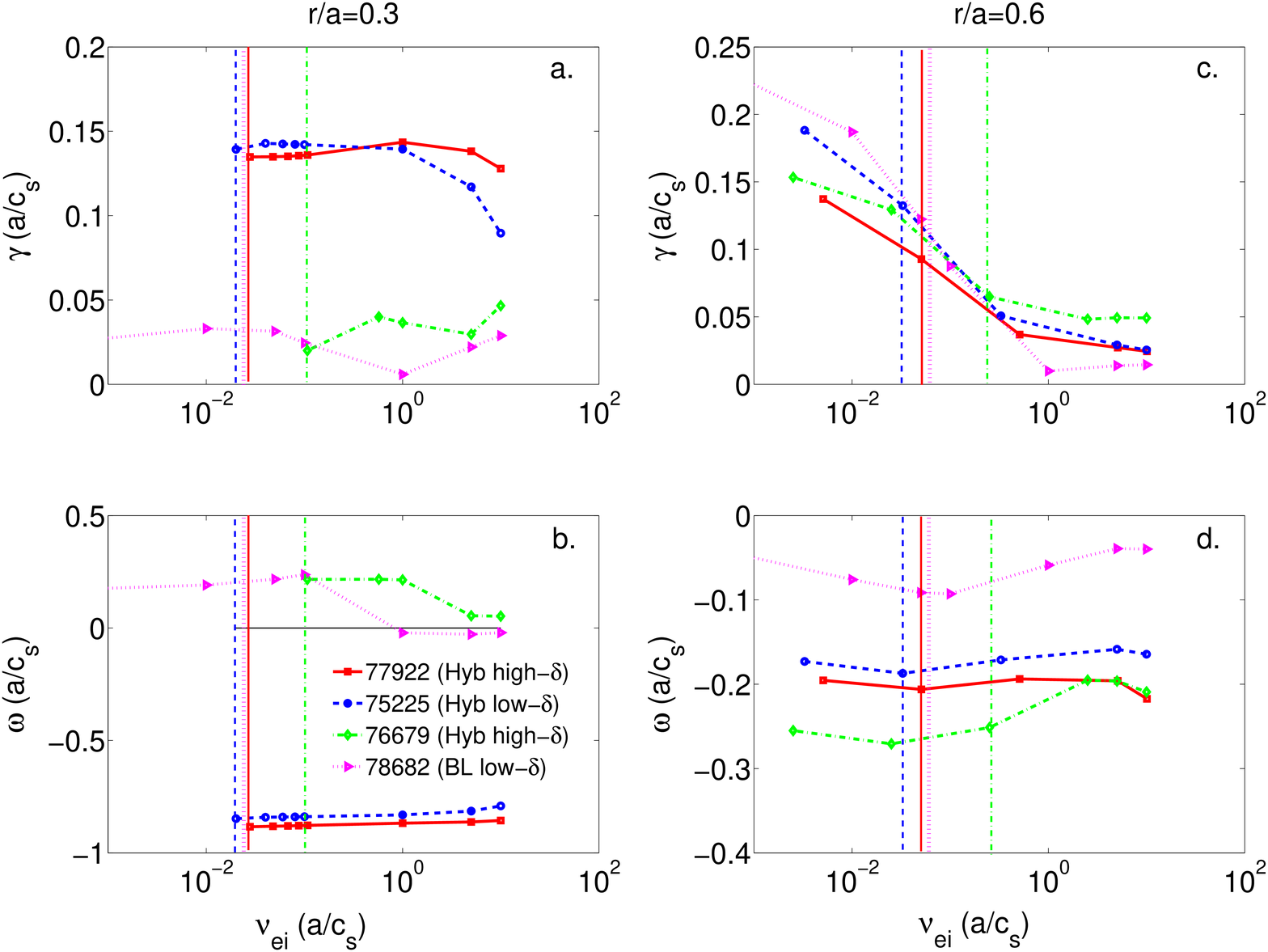}
\caption{Imaginary and real frequency of the most linearly unstable
  modes as functions of $\nu_{ei} (a/c_s)$. (a,b) at $r/a=0.3$, and
  (c,d) at $r/a=0.6$. Solid (red) lines correspond to the values for
  77922, dashed (blue) lines for 75225, dashed-dotted (green) lines
  for 76679, and dotted (mauve) lines for 78682 discharges. The
  straight lines represent the experimental values for $\nu_{ei}
  (a/c_s)$ colour coded similarly to the corresponding curves.}
\label{fig8}
\end{center}
\end{figure}

As was discussed in the previous section, collisionality between the
hybrid and baseline plasmas are usually very different since baseline
plasmas usually have higher electron density.  Therefore, we have
examined the role of collisionality in the stability properties of
these plasmas with respect to micro-instabilities.  Figures
\ref{fig8}(a-d) show the results of the collisionality scans for the two
considered radii. Note that the plots are log-linear. At the inner
core radius $r/a=0.3$, the KBMs in hybrid plasmas show very little
dependence to $\nu_{ei}$, and are stabilized only at the very high
$\nu_{ei}$, however they remain KBM. The TEM mode in high-$\delta$
baseline plasma (green dashed-dotted line) also varies little with
increase of collisionality, but MTM in low-$\delta$ baseline plasma
(mauve dotted line) is unstable for very low $\nu_{ei}$ below its
experimental value (represented by the mauve straight line) and shows
a non-monotonic trend as $\nu_{ei}$ is increased, with the maximum
around the experimental level. As was reported previously in the works
\cite{smoradipop2012,DoerkPoP2012} the non-monotonic dependence of the
MTM growth rate on collisionality is due to the fact that, on the one
hand magnetic reconnection favours finite collisionality
(although unstable MTMs have been found even in essentially collisionless cases 
\cite{DoerkPoP2012,carmodypop2013}), on the other hand, in a strong
  rate of scattering of electrons prevents the formation of a current
  layer, hence MTMs are stabilized at sufficiently high
  collisionality.

At the outer core radius $r/a=0.6$, the ITGs in all the selected
discharges are stabilized strongly as the collisionality is increased,
as seen in Figs. \ref{fig8}(c) and (d). The reason for this strong
stabilization is a collisional reduction of the non-adiabatic trapped
electron response, as the destabilizing effect of the trapped
electrons on ITG is well known, see Ref. \cite{romanelipop2007} . The
trapped electron response is mostly eliminated at
$\nu_{ei}/|\omega|\gg 1$, accordingly we observe that the growth rates
level off around $\nu_{ei}\sim c_s/a$.  The real part of the ITG mode
frequency is only weakly affected by the collisions.

In summary, the results show that, in the hybrid plasmas, the most
unstable modes regardless of their type, have higher growth rates than
in the baseline plasmas. This is not surprising since gradients and
therefore the drives of the KBM/ITG modes are higher, and
collisionality which has stabilizing impact is lower in the hybrid
plasmas, see tables \ref{table1}-\ref{table4}. However,
experimentally, the hybrids show better core confinement than baseline
plasmas, which would suggest presence of a lower level of turbulent
transport. Thus, we expect that the experimental cases correspond to
stable or only marginally unstable situations. Therefore the
resolution of this apparent contradiction requires additional physics
to be considered as will be shown below, and there is a reason to
believe that the set of geometry and profile parameters used here may
be somewhat different from the actual experimental ones, and a strong
sensitivity for some of these parameters may account for the stability
of the experimental cases. In the following section we will examine
the role of possible stabilizing mechanisms in the selected plasmas.

\subsection{Role of fast ions, safety factor and magnetic shear on the stability of the modes}

\begin{figure}[htbp]
\begin{center}
 \includegraphics[width=0.45\textwidth]{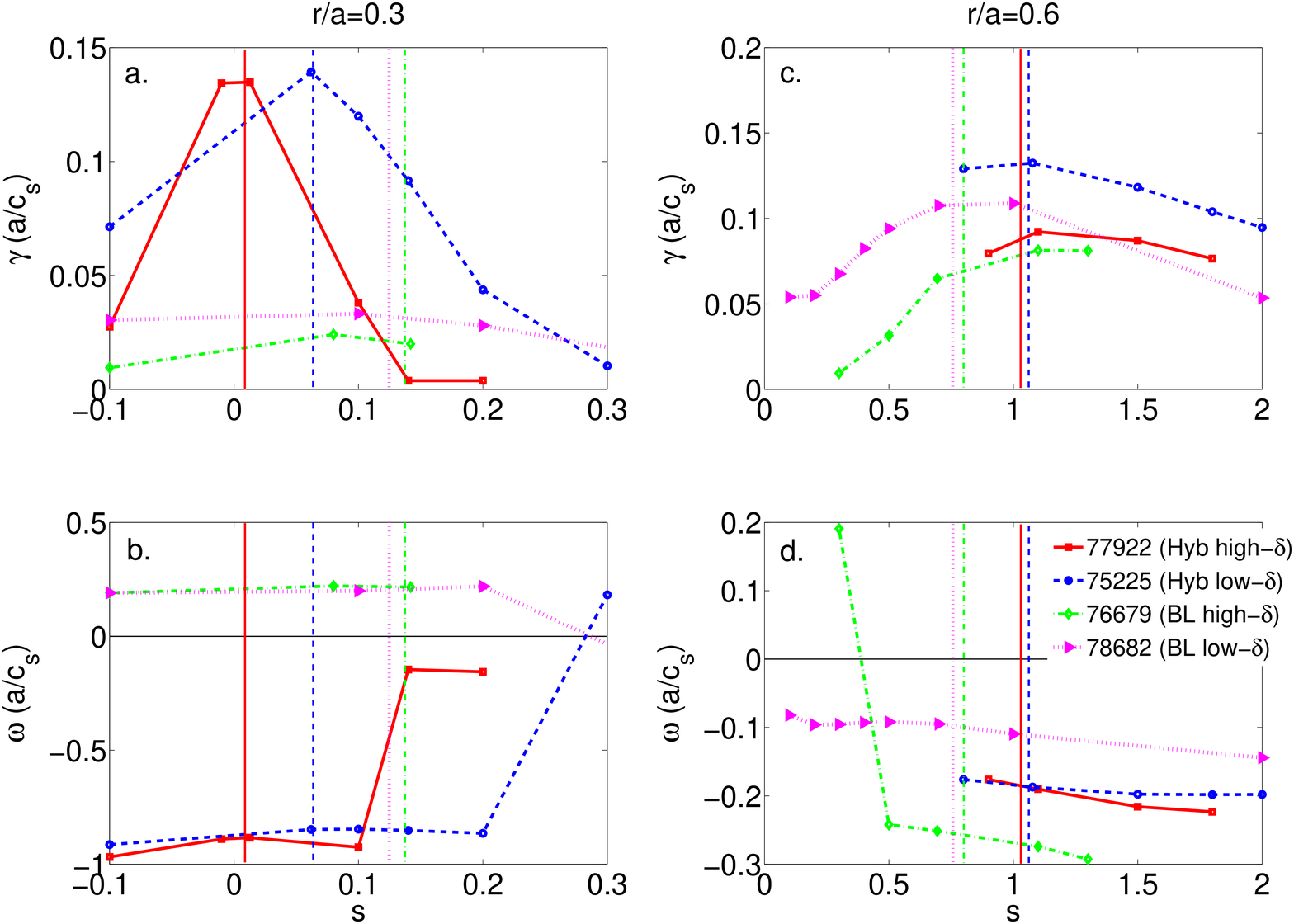} \includegraphics[width=0.45\textwidth]{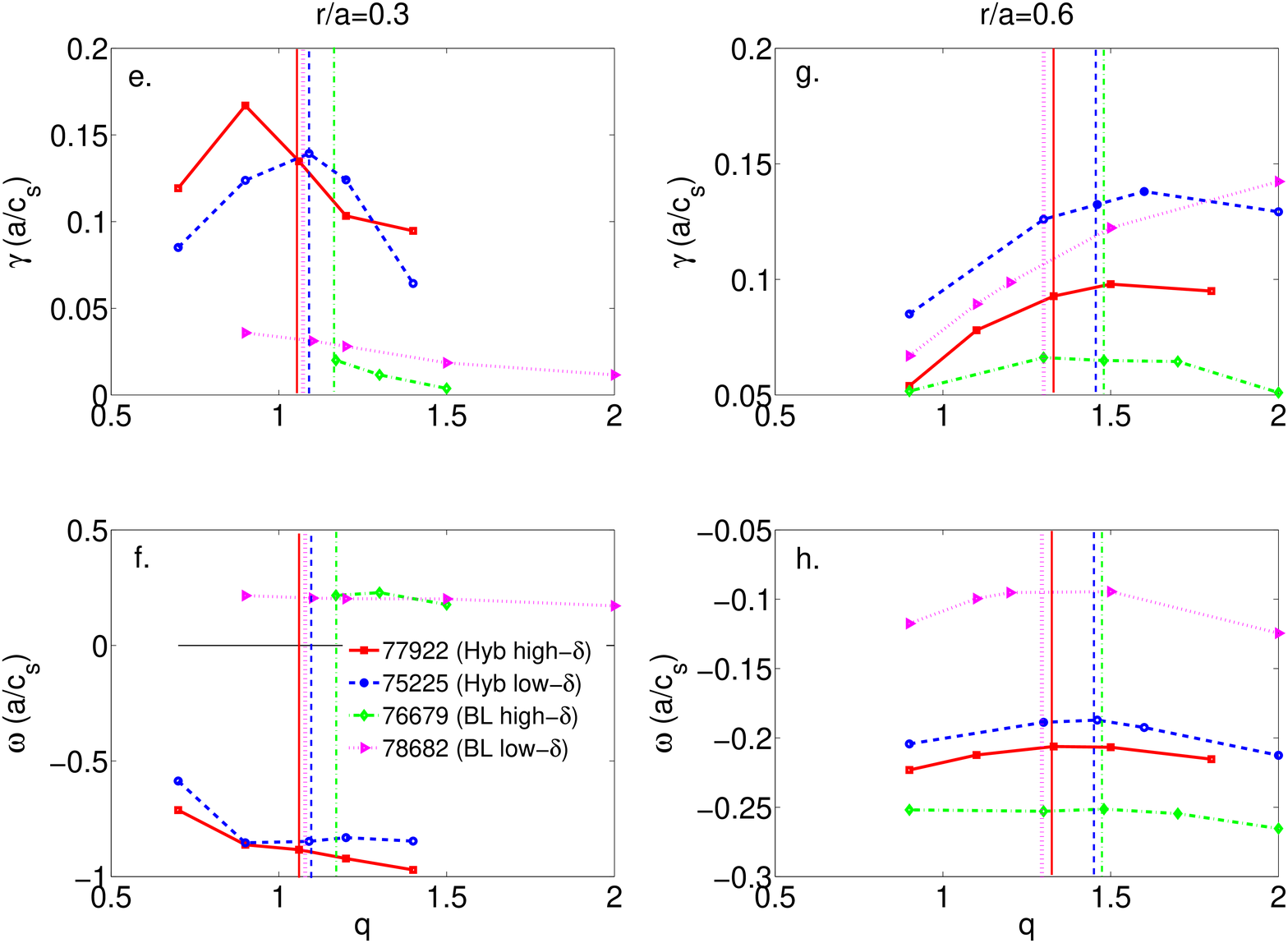}
\caption{Imaginary and real frequency of the most linearly unstable
  modes as functions of $s$ (left), and $q$ (right). (a,b) and (e,f)
  at $r/a=0.3$, (c,d) and (g,h) at $r/a=0.6$. Solid (red) lines
  correspond to the values for 77922, dashed (blue) lines for 75225,
  dashed-dotted (green) lines for 76679, and dotted (mauve) lines for
  78682 discharges. The straight lines represent the experimental
  values for $s,q$ colour coded similarly to the corresponding
  curves..}
\label{fig9}
\end{center}
\end{figure}

\begin{figure}[htbp]
\begin{center}
 \includegraphics[width=0.45\textwidth]{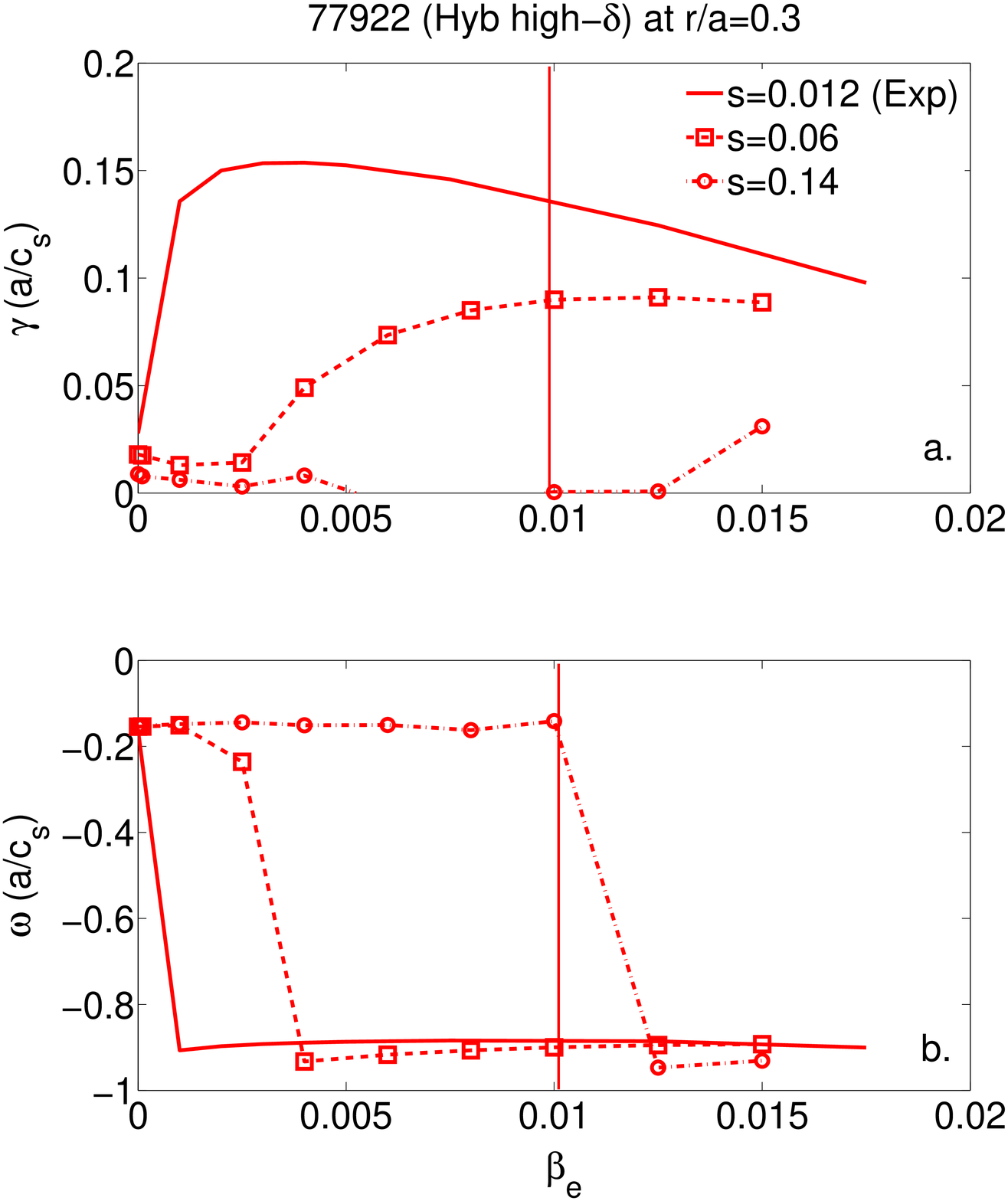} \includegraphics[width=0.45\textwidth]{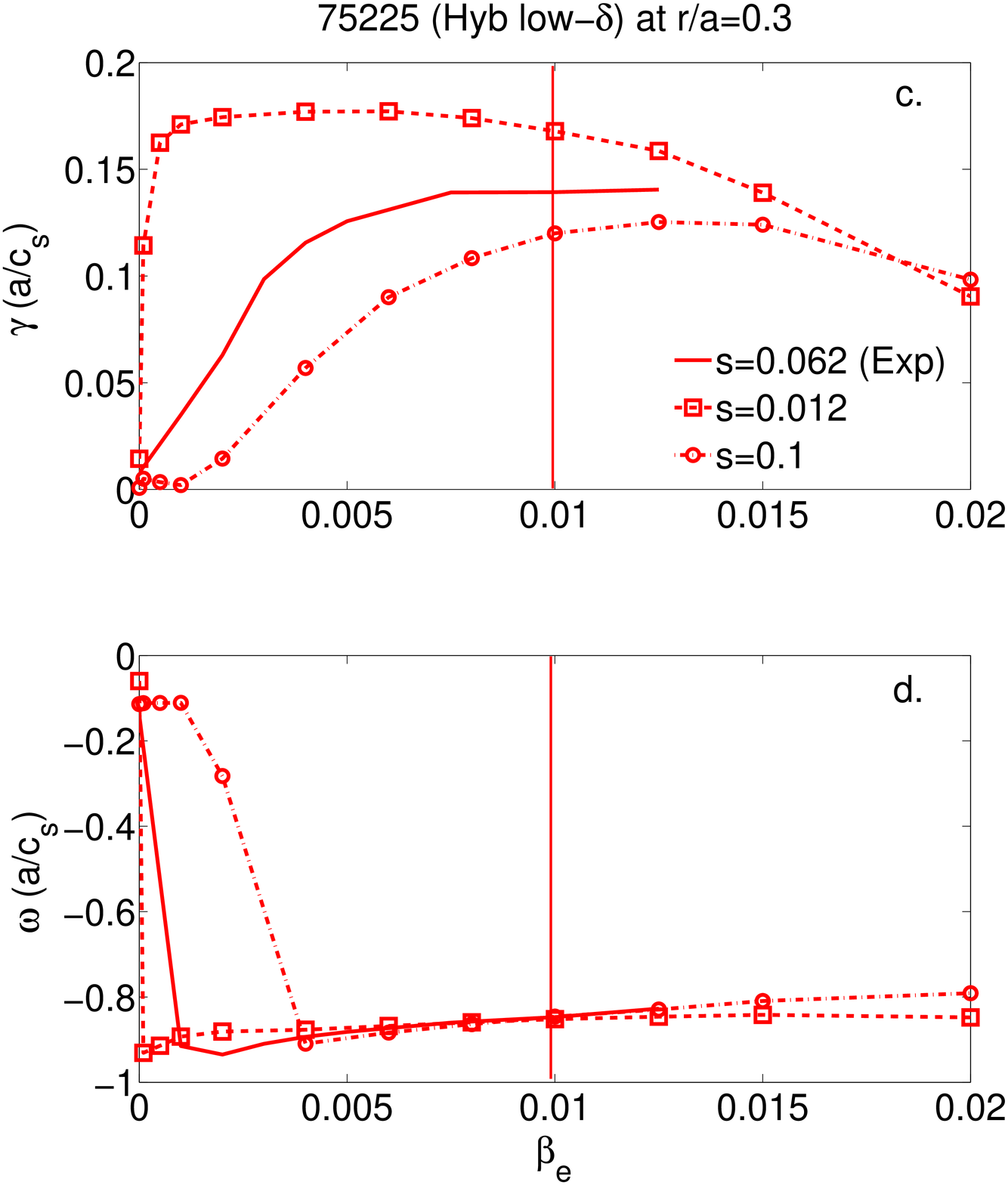}
\caption{Imaginary and real frequency of the most linearly unstable
  modes at $r/a=0.3$ as functions of $\beta_e$ for various magnetic
  shear, $s$. (a,b) show the results for 77922, and (c,d) for 75225
  JET discharges, where solid (red) lines correspond to the results
  obtained with experimental values of $s$. The straight lines
  represent the experimental values for $\beta_e$.}
\label{fig9-1}
\end{center}
\end{figure}

Figures \ref{fig9}(a-h) show the results of magnetic shear $s$ and
safety factor $q$ scans for the selected discharges at both radii. As
seen in Figs. \ref{fig9}(a,b) and (e,f) for the inner core region
($r/a=0.3$), variation in $s$ and $q$ have significant impact on the
stability of the KBMs in the hybrids while they have only minimal
effect on stability of TEM and MTM in the baseline plasmas. A
non-monotonic variation with $q$ is observed for KBM growth rates as
shown in Figs. \ref{fig9}(e,f), with the peak being around the
experimental values (shown with vertical bars). However within the
range of $q$ variation considered here, KBM remains the dominant
instability with decreased growth rate as $q$ is increased. As
illustrated in Figs. \ref{fig9}(a,b), the low magnetic shear in hybrids
($s\sim 0.01$) is a reason for the destabilization of the KBMs since
by increasing the shear \red{towards} the baseline values ($s\sim 0.1$) the
KBMs are completely stabilized and the most unstable mode would switch to
ITG type in both hybrid plasmas. The importance of low magnetic shear
in destabilization of the KBMs is further investigated by a series of
$\beta_e$ scans at various $s$, for the two hybrid plasmas, presented
in Figs. \ref{fig9-1}(a-d). As shown here, the very low magnetic shear
in hybrids results in the observed low $\beta_e$ threshold of KBMs in
these plasmas, and as $s$ is increased the threshold is pushed towards
higher $\beta_e$. Hence, it is not surprising that although baselines
plasmas have similar experimental $\beta_e$ (see tables \ref{table1})
to those of hybrids, because of their higher magnetic shear, they are
KBM stable. This suggests that from inner core instability point of
view an important parameter in hybrid plasmas is the low magnetic
shear. 

\new{We would like to note that, in a local simulation at very small shear the 
radial extent of the box size (based on the distance of resonant surfaces) gets very
  wide in terms of real radius, and ``flattening'' the profiles might
  lead to differences as compared to considering the actual profile
  variations. Nevertheless, it is often the case that the results
  remain qualitatively similar [sometimes even quantitatively, 
  as demonstrated for instance in Refs. \cite{CandyPoP2004}]. 
  This is the case in our low shear core plasmas with KBM instability, 
  as supported by global simulations (see section \ref{secglob}).}

As seen in Figs. \ref{fig9}(c,d) and (g,h), for the outer core radii
($r/a=0.6$), an increase in either magnetic shear or safety factor
will result in stabilization of the ITG modes in all selected
plasmas. However, with variations in a realistic uncertainty
  range, the ITG mode remains the dominant instability and is not
completely stabilized.

An important stabilizing parameter for the core micro-instabilities,
is the pressure gradient, $\beta^{'}\sim dp/dr$, or $\alpha_{MHD}$
parameter. It has been shown previously that in the so-called
$s-\alpha$ equilibrium which is valid at high $\beta$ and very
  large aspect ratio for circular concentric magnetic surfaces; the
high pressure gradient reduces the magnetic drift drive
responsible for the interchange instability, see
Ref. \cite{bourdelle2003}. However, in the case of the Miller
equilibrium considered here, the effect of the pressure gradient on
the drift velocity is complicated, see Ref. \cite{candy09}. The increase in pressure
gradient can be the result of the presence of a minority of energetic
ions, and previous works have shown their importance on the
suppression of ITG modes driven by gradients of the thermal population
(deuterium) \cite{romanelli2010}. Also, the impact of fast ion
dilution of the main ion species have been reported in
Refs. \cite{tardini,holland}, and recent nonlinear electromagnetic
simulations of JET experimental discharges have shown strong
stabilization by supra-thermal pressure gradients to be a key factor
in reducing tokamak ITG instabilities~\cite{jonathanPRL2013}.

\begin{figure}[htbp]
\begin{center}
 \includegraphics[width=0.8\textwidth]{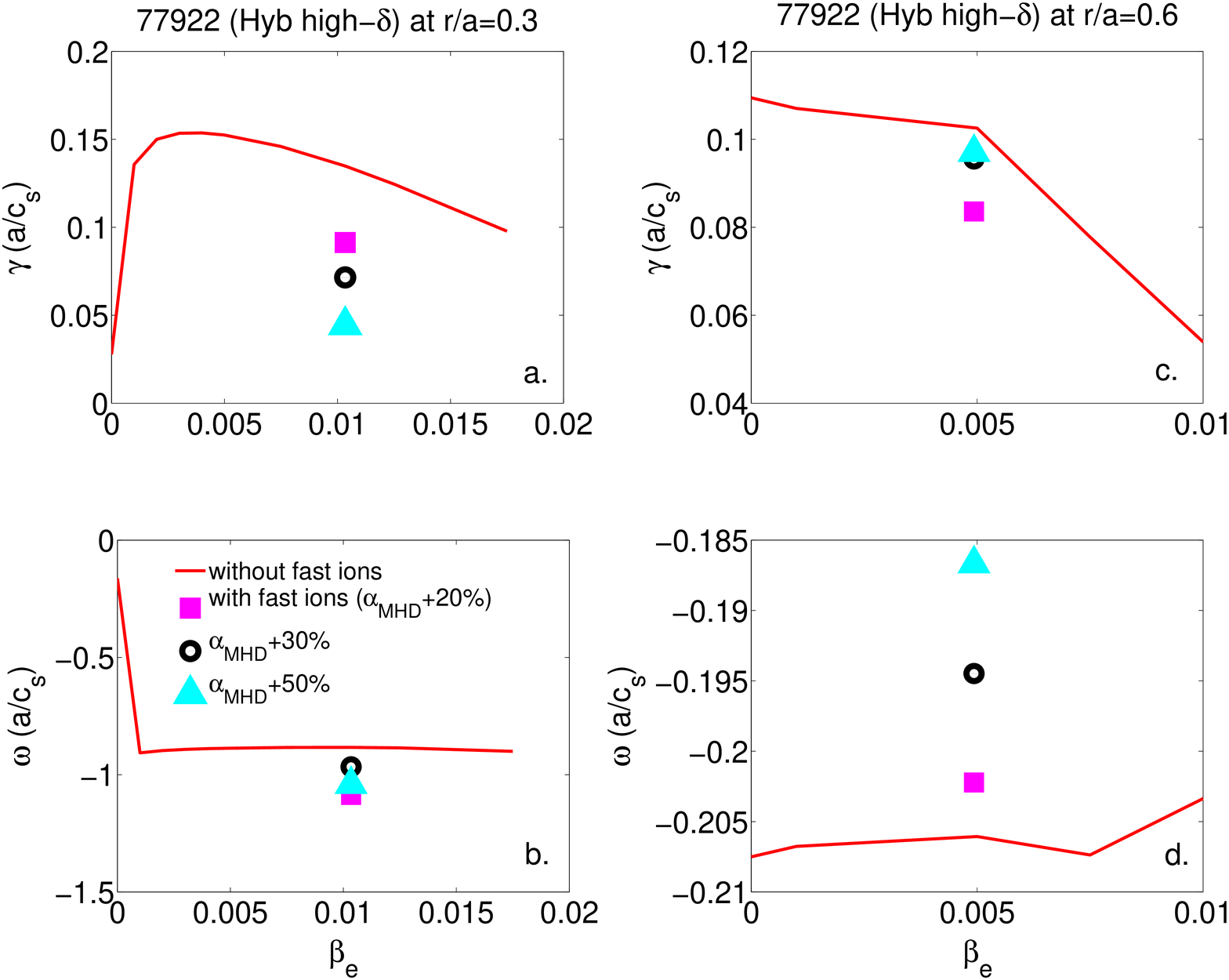}
\caption{Imaginary and real frequency of the most unstable modes for
  JET discharge 77922 as functions of $\beta_e$, where contributions
  from fast ions are included. (a,b) at $r/a=0.3$ and (c,d) at
  $r/a=0.6$. In this figure the red solid lines represent the results
  without the contribution from the fast ions, and square (mauve)
  symbols represent the results where the contribution from fast ions
  are considered. Further increase (artificially) of the impact of the
  fast ions on the $\alpha_{MHD}$ using the multiplier parameter $c_p$
  as defined in Eq. \ref{alpha}, are shown with black circle
  ($\alpha_{MHD}+30\%$) and blue triangle ($\alpha_{MHD}+50\%$)
  symbols. }
\label{fig10}
\end{center}
\end{figure}

Therefore, we have examined the impact of the fast ion population
on micro-instabilities in the core of the hybrid high-$\delta$
discharge ($77922$), by considering the fast ions as a new ion species
and using their density and temperature profiles calculated by
TRANSP. The increase in $\alpha_{MHD}$ parameter, due to fast ion
population, is calculated to be around $20\%$, see
Figs. \ref{fig10}(a-d). In this figure, the red solid lines represent
the results without the contribution from the fast ions, and square
(mauve) symbols represent the results where the contribution from fast
ions have been included. As seen here at $r/a=0.3$, a $20\%$ increase
in $\alpha_{MHD}$ significantly suppresses the KBM however, this
increase is not enough to completely stabilize the mode. At $r/a=0.6$,
however, this increase has no significant impact on the stability of
the ITG mode.

\begin{figure}[htbp]
\begin{center}
 \includegraphics[width=0.45\textwidth]{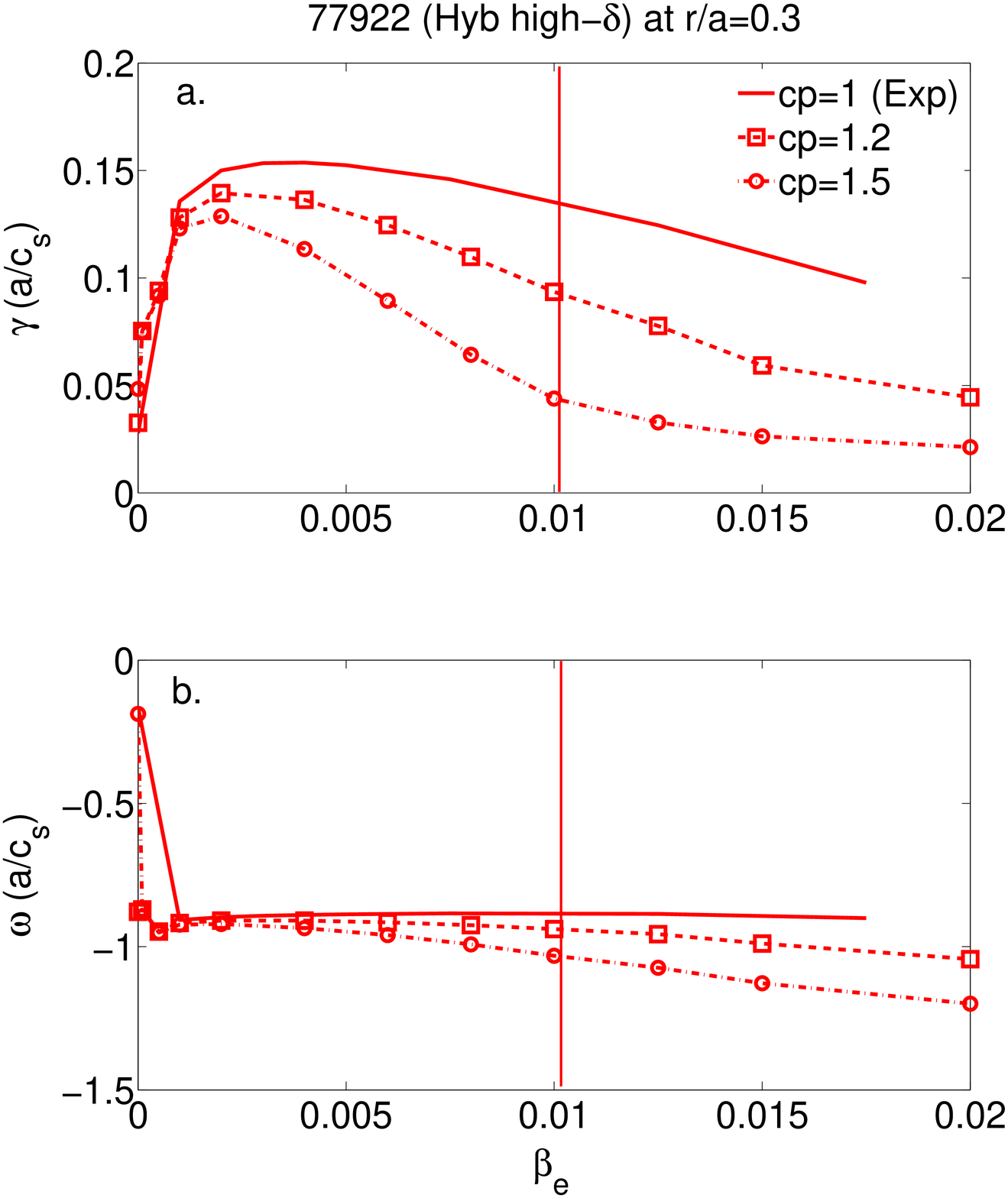} \includegraphics[width=0.45\textwidth]{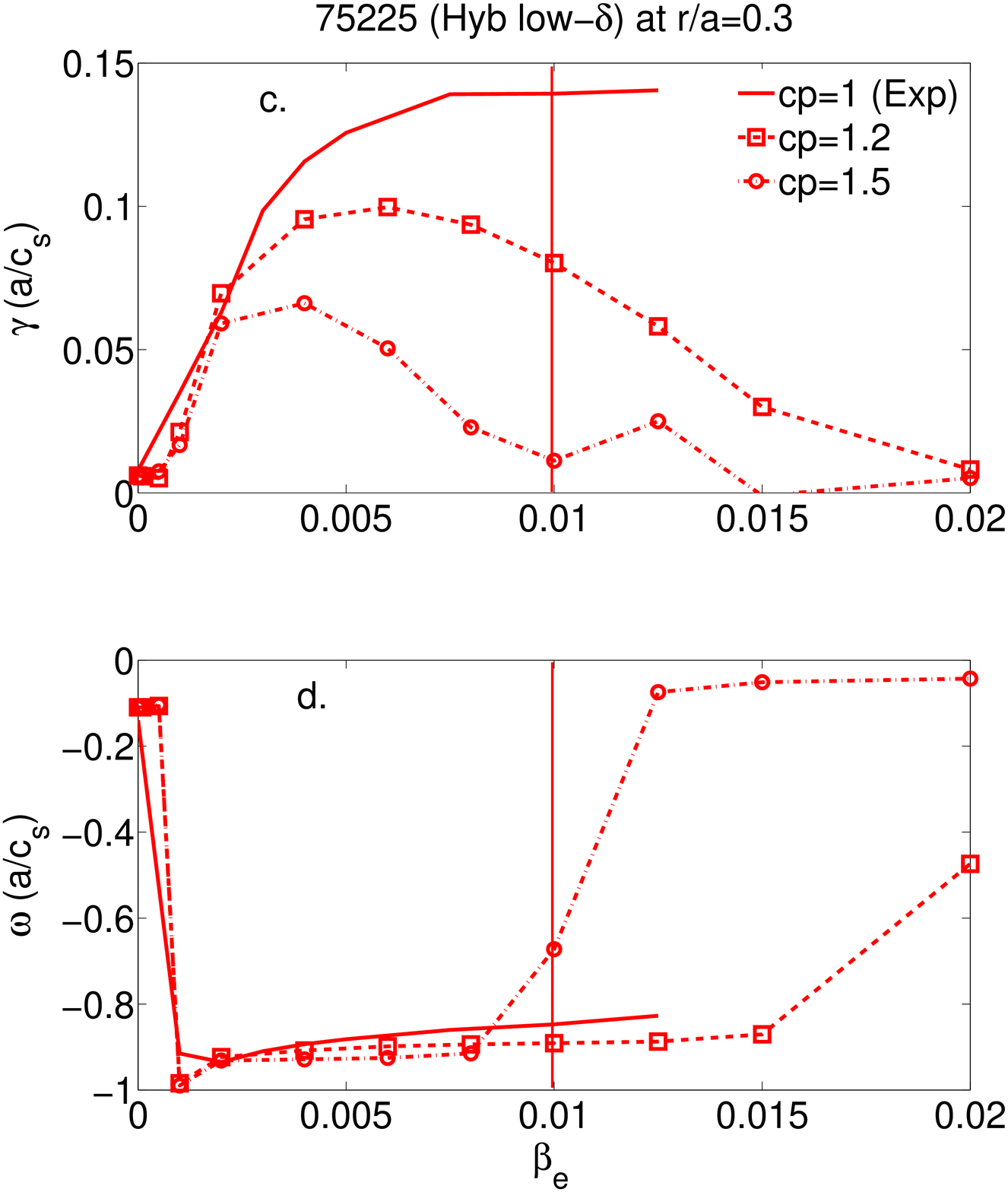}\\\includegraphics[width=0.45\textwidth]{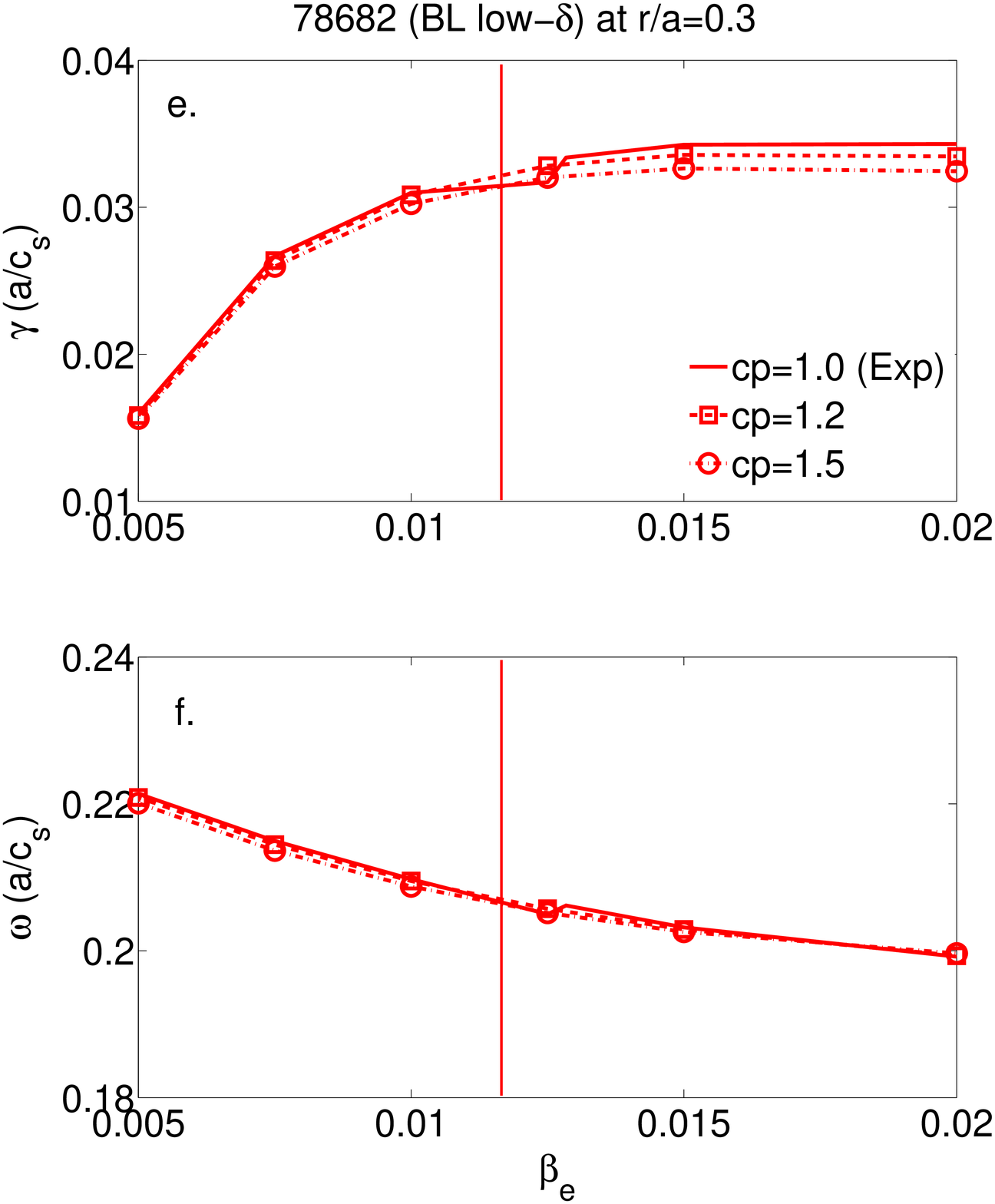}
\caption{Imaginary and real frequency of the most linearly unstable
  modes at $r/a=0.3$ as functions of $\beta_e$ for various values of
  $c_p$. (a,b) show the results for 77922, (c,d) for 75225, and (e,f)
  for 78682 JET discharges, where solid (red) lines correspond to the
  results obtained with $c_p=1$. The straight lines represent the
  experimental values for $\beta_e$.}
\label{fig10-1}
\end{center}
\end{figure}

In order to examine the sensitivity of these results to the fast ion
population and therefore, on the $\alpha_{MHD}$, we have artificially
increased $\alpha_{MHD}$ using the multiplier parameter $c_p$ as
defined in Eq. \ref{alpha}. The results are shown in
Figs. \ref{fig10}(a-d) with black circle ($\alpha_{MHD}+30\%$) and blue
triangular ($\alpha_{MHD}+50\%$) symbols. Again, these increases seem
to significantly stabilize the KBM, but their impacts on ITG mode are
minimal. These results indicate that the impact of the fast ions on
the stabilization of the micro-instabilities is more relevant at
the inner core radii, and is not expected to impact the modes
further out. This is more clearly shown in Figs. \ref{fig10-1}(a-f),
where the results of $\beta_e$ scans at various $c_p$ values for three
of the selected plasmas at the $r/a=0.3$ are illustrated. For both
hybrid plasmas, an increase in $c_p$ results in a very strong
suppression of the KBMs, and even a change to ITG mode seen in case of
low-$\delta$ hybrid. However, the lower $\beta_e$ threshold of
the KBMs do not vary much as $c_p$ is increased. In the case of
low-$\delta$ baseline plasma, the increase in $c_p$ \red{does} not
result in a significant suppression of the MTM, see Figs. \ref{fig10-1}(e) and
(f).

\clearpage 
\subsubsection{\new{Global linear scans}}
\label{secglob}
\begin{figure}[htbp]
\begin{center}
 \includegraphics[width=0.45\textwidth]{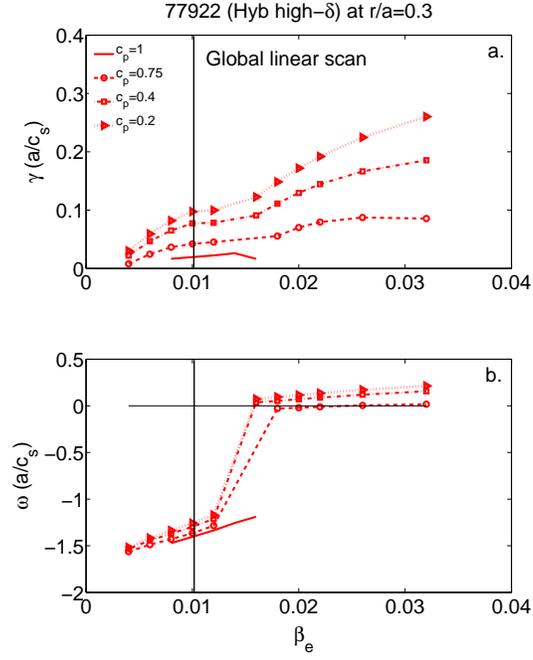}
\caption{\new{Imaginary (a) and real frequency (b) of the most
    unstable modes for JET discharge 77922 as functions of $\beta_e$,
    for various values of $c_p$. In this figure the solid (red) lines
    correspond to the results obtained with $c_p=1$. The
    \blue{vertical} line corresponds to the experimental value
    \blue{of} $\beta_e$.}}
\label{fig11}
\end{center}
\end{figure}

\new{To examine the importance of profile variations on the stability
  of KBMs found in the core of the hybrids, we have performed global
  \blue{(i.e. non-periodic boundary condition, using experimental
    profiles)} linear scans for 77922 discharge\blue{.  The simulation
    covers the radial range $r/a=0.21-0.39$ ($r/a=0.16-0.44$ together
    with the buffer zones), 46 radial grid points are used (the radial
    resolution spans the range $k_x\rho_s=0.08-0.92$), and the
    toroidal mode number is 32 corresponding to $k_y\rho_s=0.4$ at the
    reference radius $r/a=0.3$, otherwise the resolution parameters
    are identical to those for the local simulations. Carbon impurity
    is included, both electron-ion and ion-ion collisions are retained
    and $Z_{eff}$ is calculated consistently to the experimental
    profile variations.  Compressional magnetic perturbations and
    rotation are not considered.  The results of $\beta$ scans} are
  shown in Fig.~\ref{fig11}(a and b). \blue{In a global simulation the
    radial $\beta_e$ is computed from the plasma parameter profiles,
    which we scale by different constant factors to obtain
    Fig.~\ref{fig11}; the $\beta_e$ values shown refer to those at the
    reference radii (the vertical bar corresponds to the experimental
    value). The different curves illustrate the effect of the finite
    $\beta$ on the magnetic geometry (i.e. $\alpha_{MHD}$
    stabilization). The solid curve corresponds to scaling
    $\alpha_{MHD}$ consistently with the scaling of $\beta_e$, while
    besides the $\beta_e$ variation $\alpha_{MHD}$ is scaled down by
    an additional factor $c_p$ in the other curves. }}

  \new{As seen in \blue{Fig.~\ref{fig11}}, the global results confirm
    that KBMs are linearly unstable at the experimental values of
    $\beta_e$ (red solid lines). \blue{We find that retaining} global
    profile variations \blue{result} in significant reduction of the
    KBM growth rates. However, \blue{these} KBMs are \blue{found to
      be} strongly localized radially (\blue{this was also the case}
    in the local simulations shown in \blue{Fig.}~\ref{fig5} where
    eigenmodes \blue{exhibit} a very elongated structure in ballooning
    space corresponding to small radial structures). \blue{Thus,} the
    impact of profile variation in the global simulations does not
    result in their complete stabilisation\blue{,} even though the
    radial width where the circumstances support unstable \blue{KBMs
      are significantly} smaller than that of the local simulations
    (which is large due to the low magnetic shear).}

\new{As seen in Fig. \ref{fig11}(a and b) \blue{comparing the
    different curves, that $\alpha_{MHD}$} stabilization of KBMs are
  stronger in global simulations than in local simulations, \blue{thus
    a stronger impact of fast ions is expected. We note, that for
    $\beta_e$ values above the experimental value, another mode
    becomes unstable with real frequencies much closer to zero (and a
    bit more on the positive side). This other mode was not seen in
    local simulations; a reason for which is that while the KBM was
    more localized around the reference radius $r/a=0.3$, this mode is
    localized closer to the outer radial boundary, above $r/a=0.4$,
    where the local parameters are different. }}

In summary, our findings show the importance of the fast ion
contribution to the $\alpha_{MHD}$ and of the magnetic shear $s$ on
the onset and the stability of electromagnetic modes at the inner core
radius in the selected hybrid discharges. We find that flat
$q$-profiles in hybrid plasmas and therefore, the low magnetic shear
can result in a shift of the $\beta_e$ threshold of the
electromagnetic modes like KBMs towards very low $\beta_e$ in these
plasmas. However, a strong suppression of KBMs due to the presence of
fast ions can be expected for the inner core region. In the baseline
plasmas however, neither the magnetic shear or fast ion contributions
to the $\alpha_{MHD}$, seem to result in a significant suppression of
MTM/TEM modes. Thus, the good core confinement observed at $r/a=0.3$
in the selected hybrid plasmas, can perhaps be explained as a result
of a combination of these two effects: low magnetic shear and fast
ions. Ref. \cite{jonathanPRL2013} also reports that the nonlinear
electromagnetic stabilization due to fast ions is seen to be more
effective at low magnetic shear. 

\new{\blue{Looking at the intersection of the solid curve and the
    vertical bar in Fig.~\ref{fig11}a, we find that at the
    \emph{nominal} experimental values the KBM growth rate is rather
    small.} Within the experimental uncertainties therefore, we expect
  KBMs to be very close to marginality in these hybrid plasmas\blue{,
    but certainly, they are not deeply stable or strongly
    unstable. They may play the role of providing} the feedback to the
  confinement, i.e. if the profiles were just a little bit steeper, or
  $\beta$ was a bit higher, then \blue{these} KBMs would start to grow
  and reduce the profiles back towards marginality.  This is analogous
  of the often observed stiff profiles close to (the nonlinear)
  marginal stability of drift waves\blue{. An extended region of very
    low shear is essential to the existence of these modes, since
    $\beta_e$ is not too high to destabilize KBMs at $\mathcal{O}(1)$
    values magnetic shear.}}

At the outer core radius ($r/a=0.6$), the picture is different since
the ITG modes here, are not strongly stabilized by fast ion
contributions or variations of the magnetic shear and $q$. An
important stabilizing factor in this region however, is the
$\mathbf{E}\times\mathbf{B}$ flow shear, and hence, we have
investigated the impact of the flow shear in a series of local
non-linear simulations at various radii for three of the selected
discharges. The results are presented in the following section.

\clearpage 
\section{Non-Linear analysis and the impact of $\mathbf{E\times B}$}
\label{sec:nonlin}
In this section the effects of $\mathbf{E}\times\mathbf{B}$ \red{flow}
shear have been investigated in non-linear electromagnetic
gyro-kinetic turbulence simulations with GYRO code. These runs
include both shear- and compressional magnetic perturbations:
$\delta B_\perp$ and $\delta B_{\parallel}$, respectively. We use a
$128$-point velocity space discretization (8 pitch angles, 8
energies, and 2 signs of velocity). 16 toroidal mode numbers were
used. The maximum $k_{\theta}\rho_s=0.8$ 
\new{(with the grid size $\Delta k_{\theta}\rho_s=0.05$)} used in all simulations ,
except for the baseline high-$\delta$ case where the maximum
$k_{\theta}\rho_s=2.5$ \new{(with the grid size $\Delta k_{\theta}\rho_s=0.1$)} was used. 
\new{The box size was $L_x/\rho_{s}=L_y/\rho_{s}= 122$, and the time steps used for the 
KBMs at $r/a=0.3$ were $\Delta t=0.00025 (a/c_s)$ and for the ITGs at $r/a=0.6$ 
were $0.001(a/c_s)$}. The simulation time was sufficiently long to achieve 
statistically steady-state levels with adequate time stepping and box sizes. 
We have also performed convergence tests which
establish the adequacy of the time step and box size.

\begin{figure}[htbp]
\begin{center}
\includegraphics[width=0.43\textwidth]{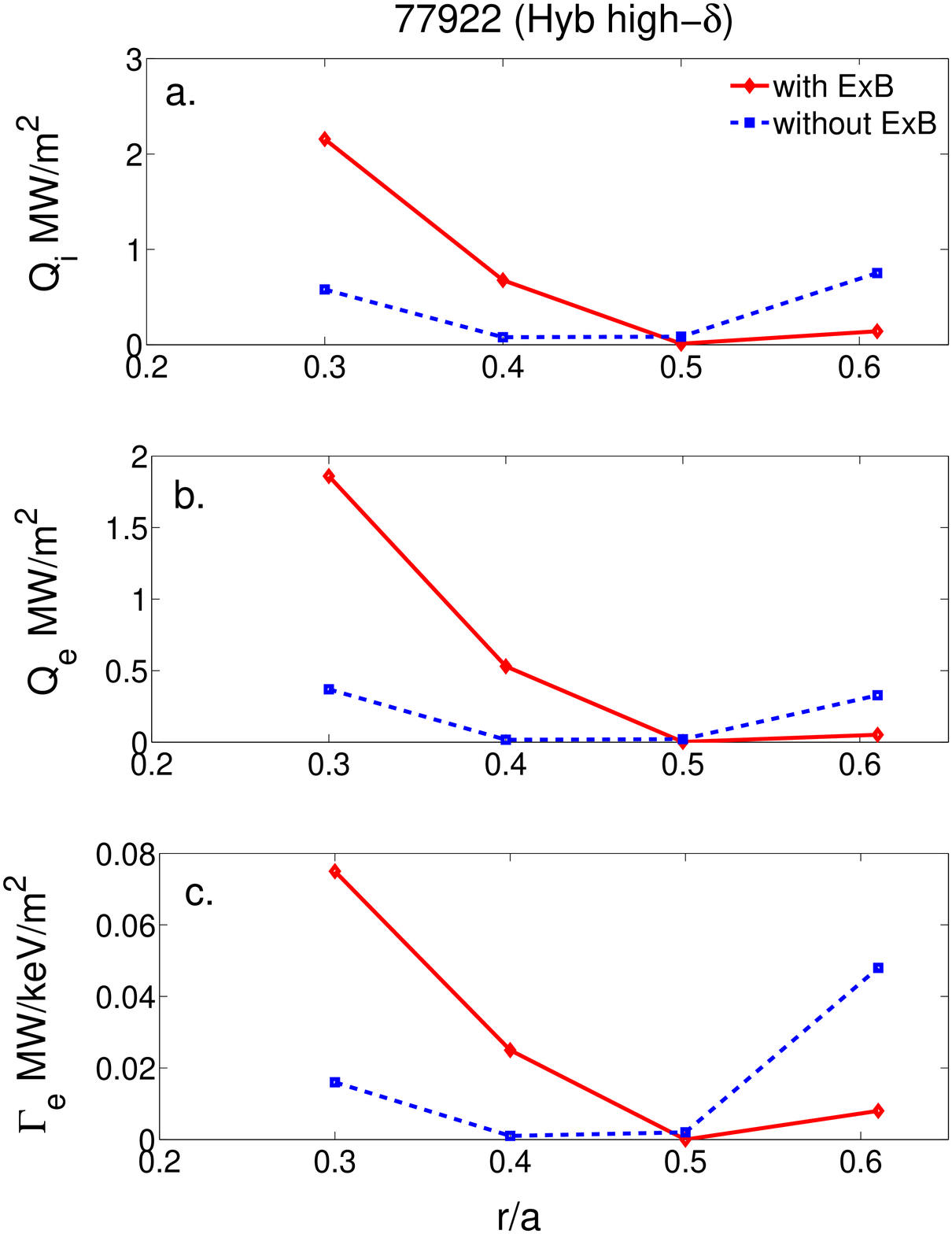}\includegraphics[width=0.43\textwidth]{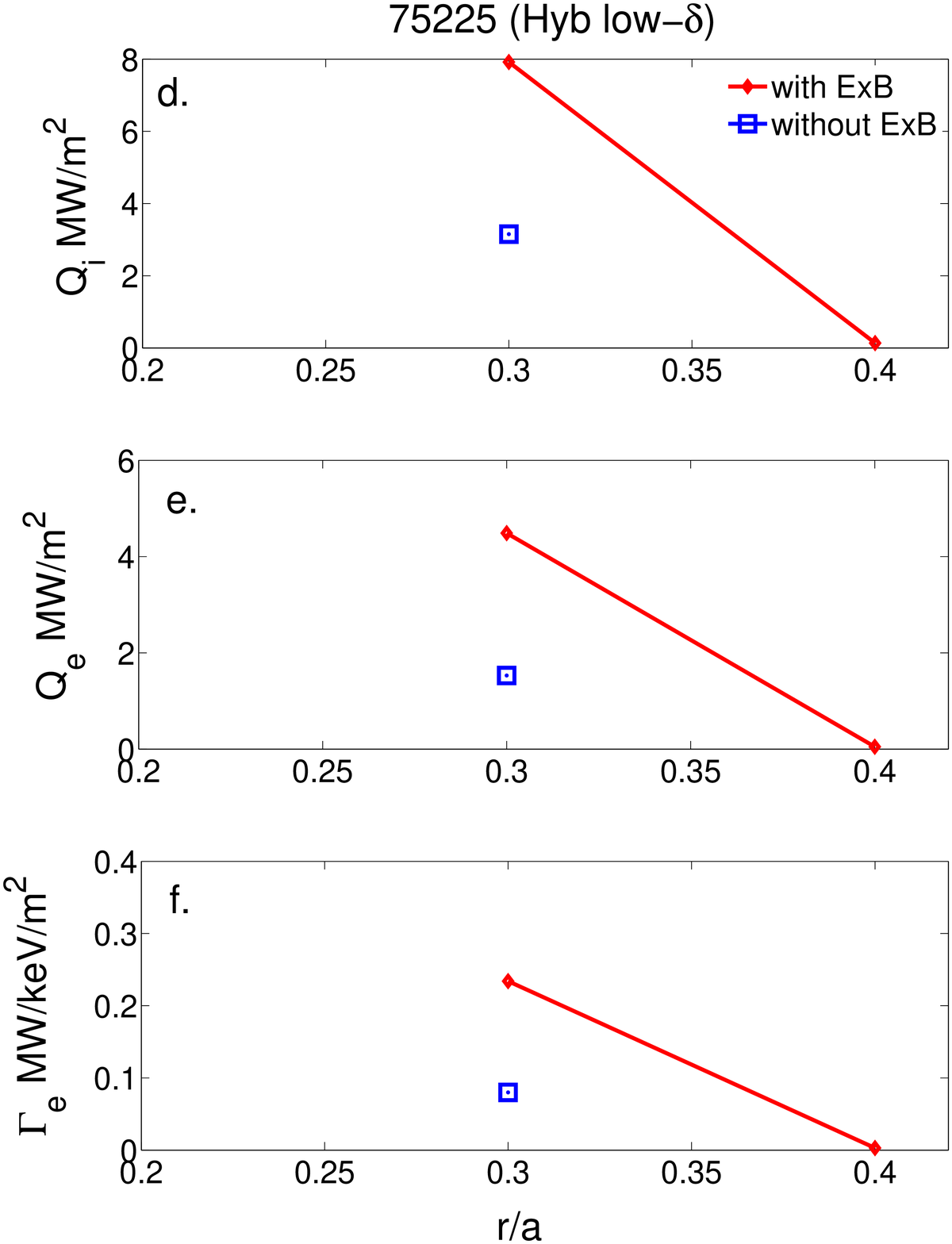}
\caption{Non-linear heat fluxes of ion and electron, and particle flux
  of electrons versus $r/a$ for the \newnew{selected hybrid discharges
  (a-c) 77922, (d-f) 75225.} Solid (red) lines
  correspond to the results obtained with and dashed (blue lines)
  without $\mathbf{E}\times\mathbf{B}$ shear effects.}
\label{fig12}
\end{center}
\end{figure}

\begin{figure}[htbp]
\begin{center}
\includegraphics[width=0.43\textwidth]{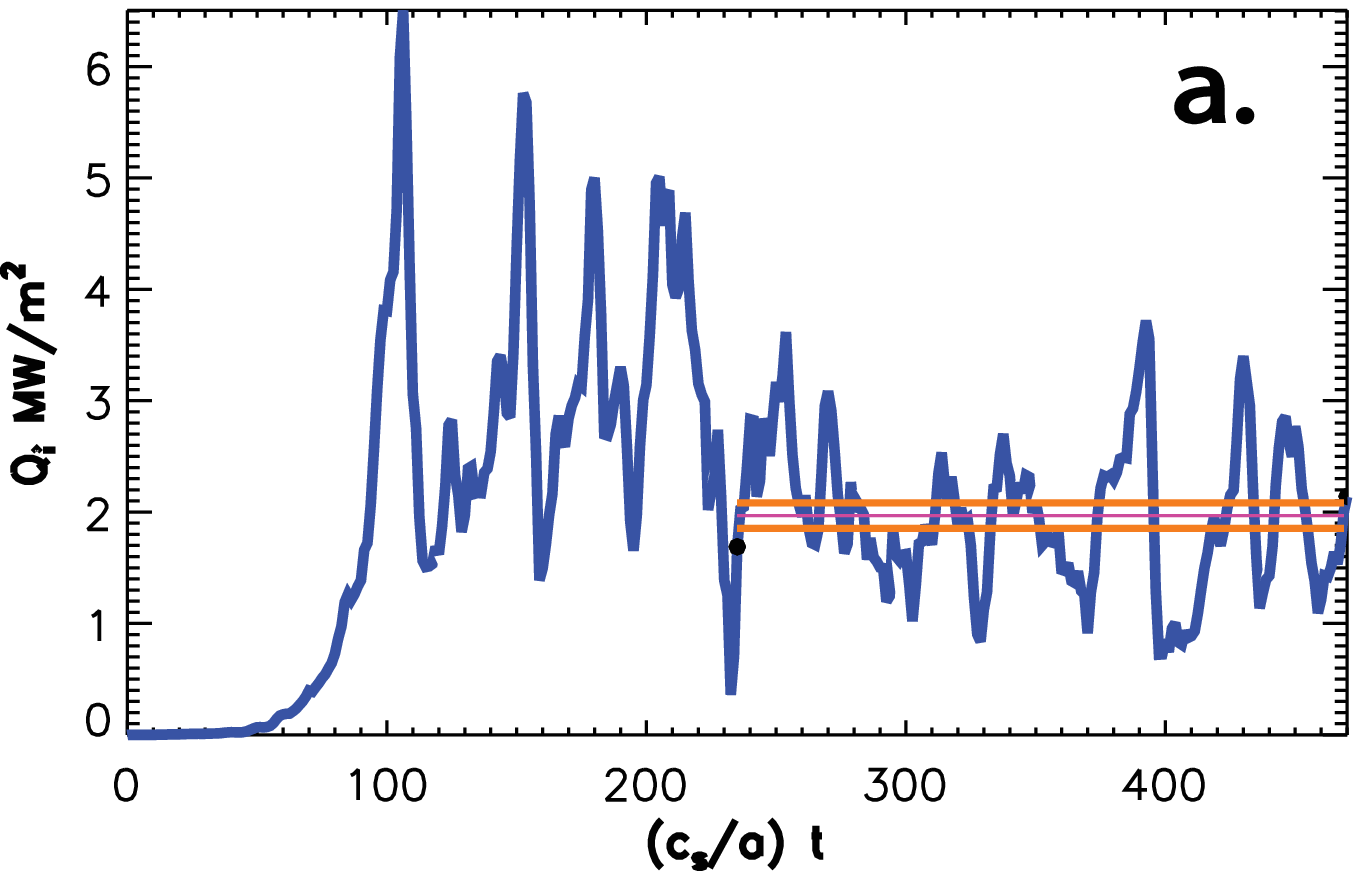}\includegraphics[width=0.43\textwidth]{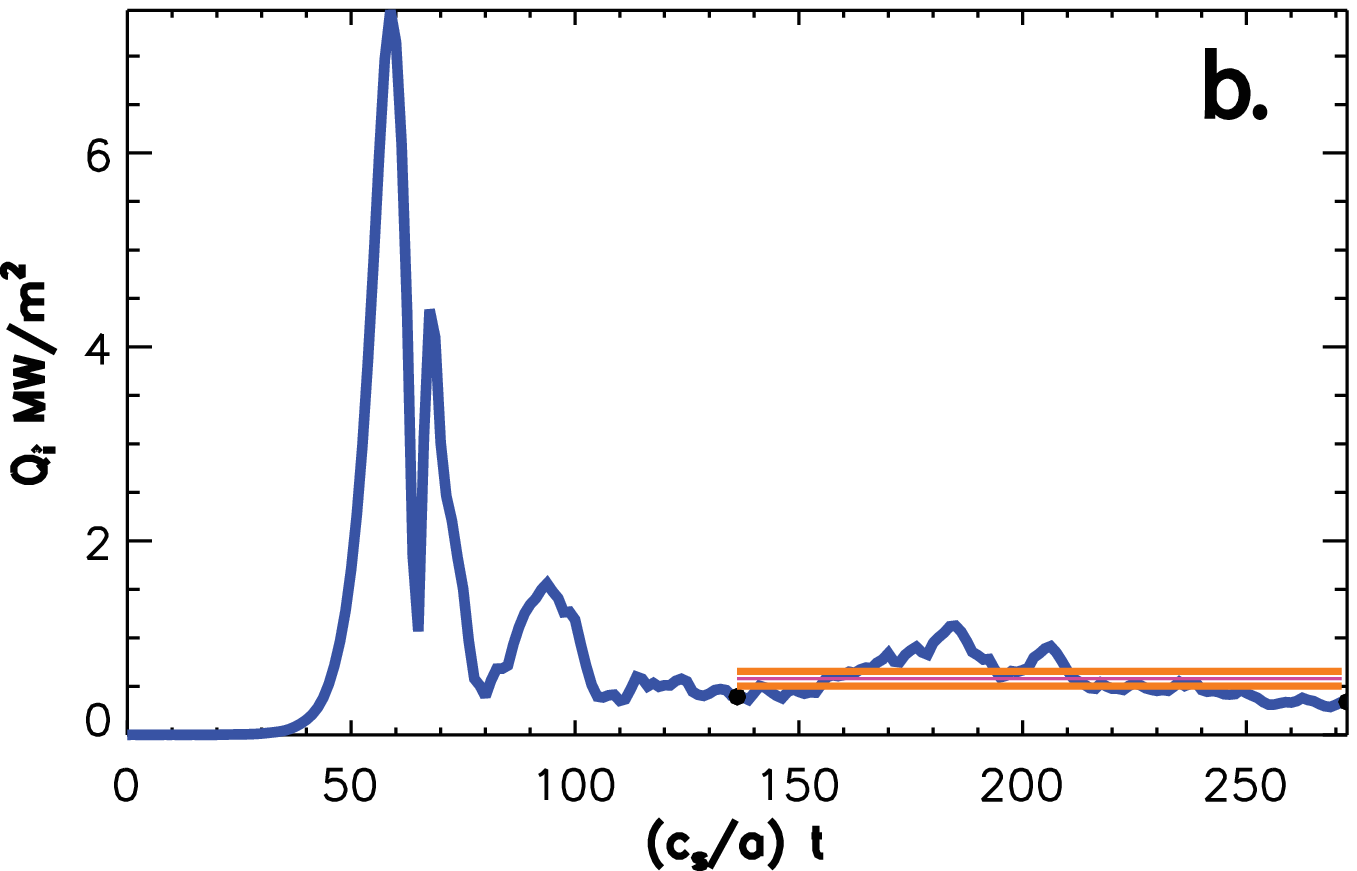}\\\includegraphics[width=0.43\textwidth]{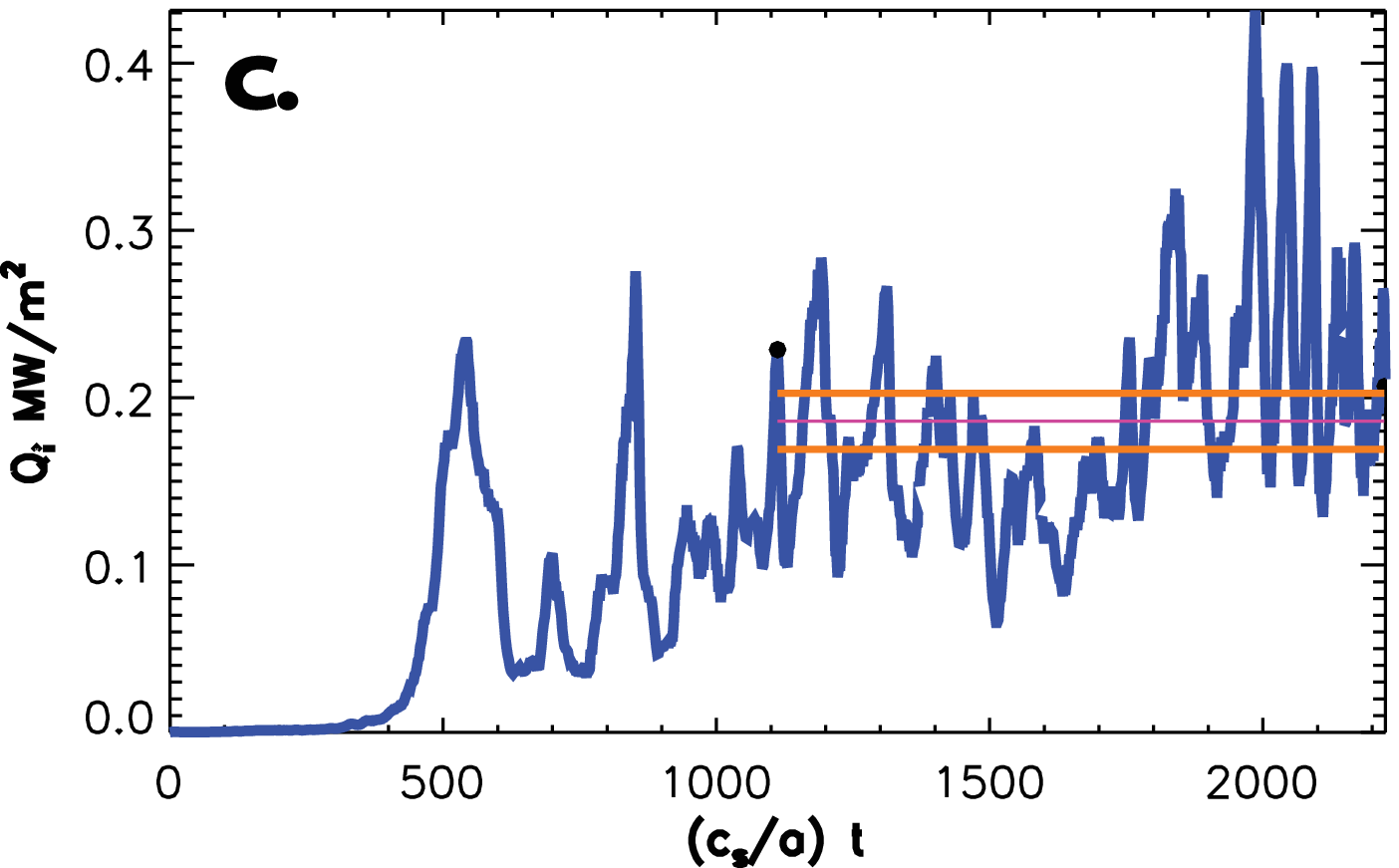}\includegraphics[width=0.43\textwidth]{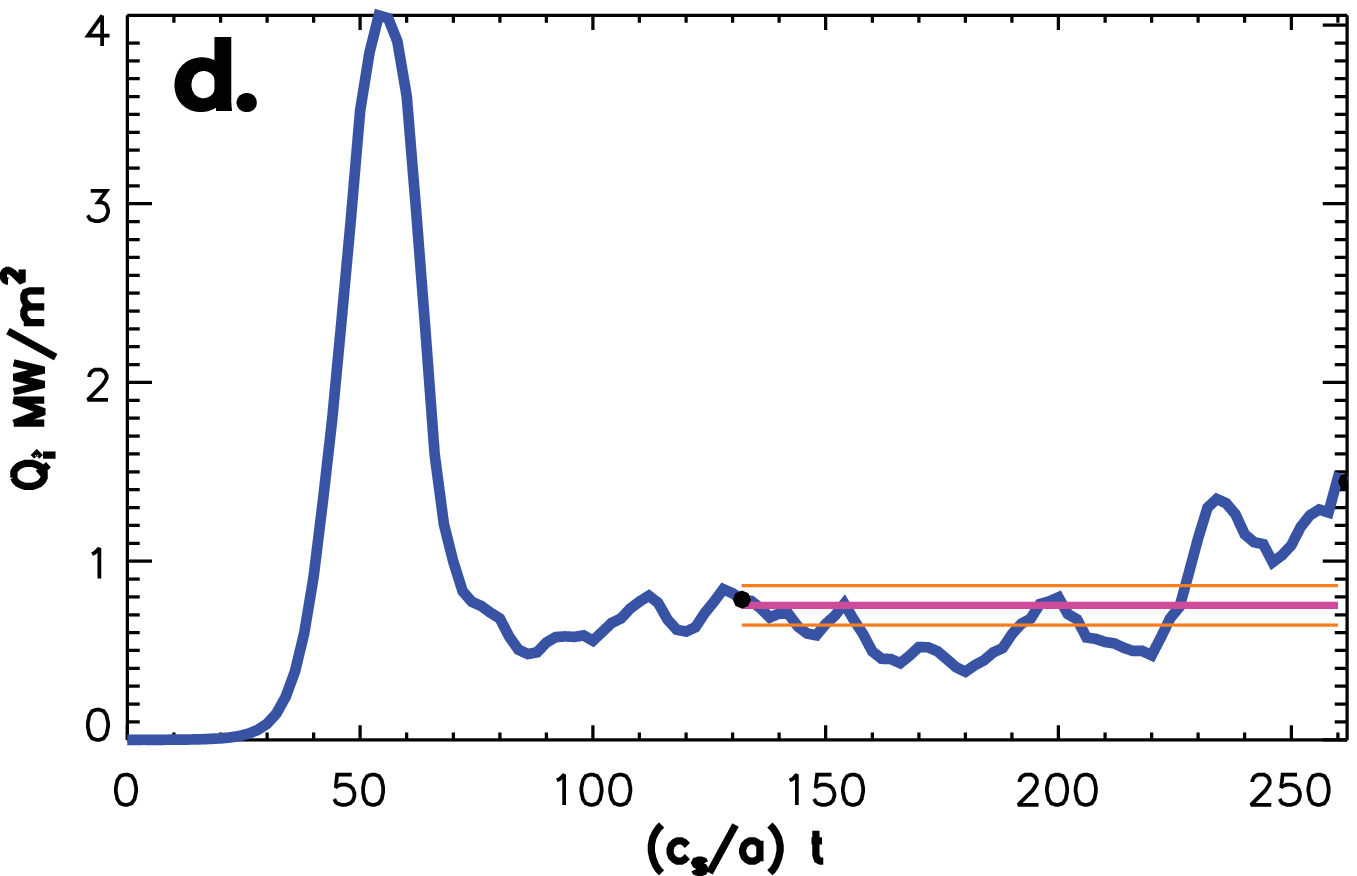}
\caption{\newnew{Time traces of the non-linear heat fluxes of ions for 77922 discharge, at $r/a=0.3$: (a) with and (b) without $\mathbf{E}\times\mathbf{B}$ shear, and at $r/a=0.6$: (c) with and (d) without $\mathbf{E}\times\mathbf{B}$ shear.}}
\label{fig13}
\end{center}
\end{figure}

The non-linear heat and particle fluxes calculated by local non-linear
GYRO simulations for \newnew{discharges 77922 and 75225 are shown in 
Figs. \ref{fig12}(a-f)}. The results are shown for cases with and without
taking into account the effect of flow shear. \new{The corresponding time 
traces of the non-linear runs for 77922 shot at $r/a=0.3, 0.6$ with and without 
$\mathbf{E}\times\mathbf{B}$ shear, are shown in Figs. \ref{fig13}(a-d).} 


As can be seen in Figs. \ref{fig12}(a-f), for $r/a<0.5$ core region, the
non-linear fluxes are significantly higher in hybrid plasmas when
flow shear effects are taken into account (red solid lines), in
comparison to the case where these effects are excluded (dashed blue
lines). The higher fluxes in these plasmas are due to destabilization
by Parallel Velocity Gradients (PVG) on the KBM/ITG modes in this
region. Using that in \gyro the effects of parallel and the
  perpendicular flow shear can be changed independently we could
investigated this effect further for the $r/a=0.3$ in hybrid
high-$\delta$ plasma by artificially taking out the contribution from
PVG. Indeed, we find that when these effects are not included,
the fluxes are similar to those obtained without considering the full
$\mathbf{E}\times\mathbf{B}$ effects. The destabilization impact of
PVG on ITG modes have been reported previously, where it has been
shown that in the presence of a large $\gamma_p$ the transport may not
be quenched by $\mathbf{E}\times\mathbf{B}$ shear, see
Refs. \cite{kinseypop2005,barnsPRL2011}. Recent studies presented in
Ref. \cite{jonathanPRL2013} also, show destabilization of the ITG
modes due to PVG in the core of JET plasmas. 

However, PVG destabilizing impact decreases with decreasing
geometrical factor $Rq/r$ \cite{highcockPRL2012}, and therefore, the
$\mathbf{E}\times\mathbf{B}$ shear can be expected to strongly
stabilizes the turbulence in the outer core region and indeed as can
be seen in Figs. \ref{fig11}(a-c), in hybrid high-$\delta$ at $r/a=0.6$
the turbulence fluxes are significantly reduced. 
\newnew{Note that the points on the Fig. \ref{fig12} with fluxes close to zero are nonlinearly stable.}

\newnew{In the case of the baseline high-$\delta$ plasma we have found that all the radii studied ($r/a=0.3, 0.4, 0.6$) were nonlinearly stable for the nominal values of plasma parameters. Therefore, we have not included simulation results for this shot.}

\begin{figure}[htbp]
\begin{center}
\includegraphics[width=0.45\textwidth]{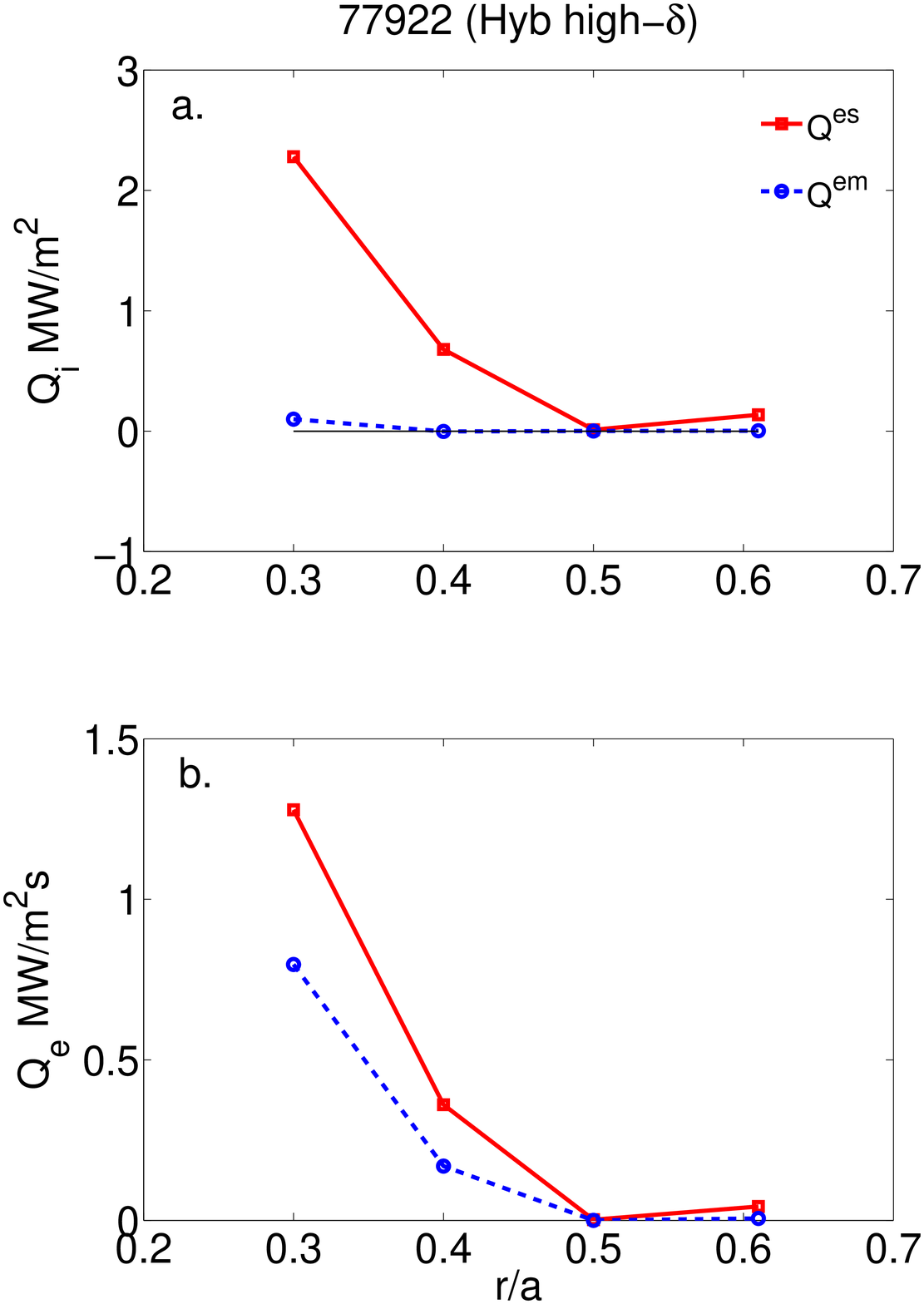}
\caption{\newnew{Non-linear ion (a) and electron (b) heat fluxes
    versus $r/a$ for the 77922 discharge. Solid (red) lines correspond
    to the electrostatic ($Q^{es}$) component of the flux and dashed
    (blue lines) correspond to electromagnetic ($Q^{em}$) component of
    the fluxes.}}
\label{fig14}
\end{center}
\end{figure}

\newnew{Note that if the electromagnetic components of the transport fluxes ($Q_{e,i}^{em}$) remain close to zero, as has been shown previously in Refs. \cite{candy,Pueschel08}, one can expect that the overall transport properties do not rely strongly on ``flutter'' transport and therefore, linear analyses are relevant and the relative sizes of the electrostatic transport fluxes ($Q_{e,i}^{es}$) are consistent with quasilinear estimates. However, this is not true for the core of the hybrid shots considered here where the KBMs are the most dominant unstable modes, and as can be seen in the Fig. \ref{fig14} (a,b) (where the electromagnetic and electrostatic components of the computed non-linear fluxes are shown for the shot 77922 as function of $r/a$), there is a significant contribution from the electromagnetic transport to the electron heat flux $Q_e$. This has been observed also when MTMs are unstable since the electron heat flux is mostly $A_\|$-driven transport in that case \cite{DoerkPoP2012}. It may be that these plasmas are unstable to a hybrid of KBMs and some other, electron driven modes, and further non-linear analysis are important to identify them. However, the non-linear simulation costs for these low shear KBM unstable plasmas proved to be prohibitive, and therefore we were not able to include further analysis at present. The importance of the ITG stabilisation with fast ions shown in the work of J. Citrin \cite{jcritin2014} presents encouraging results and supporting evidence that the present results are important. Therefore, we think that the main conclusion can be made from the current findings. In future works, we plan to further study the impact of fast ions on the KBMs through further non-linear simulations.} 



\section{Conclusions}
\label{sec:conclusions}
The experimental observations show that Hybrid plasmas can have
significantly improved normalized confinement \newred{with respect to
  the} $H_{98}(y,2)$ H-mode scaling. However, the underlying physics
basis for the observed increased normalized confinement remained
somewhat unclear. \newred{Recent studies support that significant part
  of the confinement improvement happens in the core of hybrid
  discharges.  Here, this hypothesis is tested in the form of a
  gyrokinetic study of two hybrid and two baseline discharges, using
  local and global linear and non-linear \gyro simulations.}


The results show that in the inner core ($r/a=0.3$) the
characteristics of the dominant unstable modes in the selected hybrid
and baseline plasmas are very different, with KBMs unstable in hybrids
and MTM/TEM modes unstable in baselines. For the outer core
($r/a=0.6$), however, ITGs are found to be the dominant instability in
all the selected discharges. In the inner core of the selected hybrid 
discharges our findings show the importance of
$\alpha_{MHD}$-stabilization of electromagnetic instabilities in the
presence of low magnetic shear. We find that, as theoretically expected, the flat $q$-profiles in
hybrid plasmas can
result in a downward shift of the $\beta_e$ threshold of KBMs. The
  global linear simulations \newred{confirm the local linear results that KBMs} are linearly unstable at
  experimental values of $\beta_e$ in the hybrids, although
  global profile variations can result in significant reduction of the
  KBM growth rates. Within the experimental uncertainties therefore,
  we expect KBMs to be very close to marginality in these hybrid
  plasmas and that they may provide the feedback to the confinement,
  i.e. if the profiles were just a little bit steeper, or $\beta$ was
  a bit higher, then KBMs would start to grow and reduce the profiles
  back towards marginality. \blue{We also find that the $\alpha_{MHD}$
    ($\sim$ fast ion)} stabilization of KBMs is stronger in global
  simulations than in local simulations. Thus, a strong suppression of
  KBMs due to the presence of fast ions can be expected for the inner
  core region.  
  
In the baseline plasmas however, neither the magnetic
shear or fast ion contributions to the $\alpha_{MHD}$, result in a
significant suppression of MTM/TEM modes. Thus, the good core
confinement observed at $r/a=0.3$ in the selected hybrid plasmas, can
perhaps be explained as a result of a combination of these two
effects: low magnetic shear and the fast ion stabilization of
KBMs. Comparing the non-linear heat fluxes computed by {\sc gyro} and
obtained by power balance analysis with TRANSP for the inner core
region show larger disagreement with higher values computed by {\sc gyro},
than for the outer region. As it was shown in
Ref. \cite{jonathanPRL2013} stabilisation of the ITG modes due to fast
ions can be even more efficient non-linearly, therefore, we suspect
that also for the KBMs found in this region, the inclusion of the fast
ions will result in a more significant stabilisation of the modes
non-linearly and better agreement with fluxes obtained by TRANSP.

 
In the outer core ($r/a=0.6$), the picture is different, since the ITG
modes there are not strongly stabilized by fast ion contributions,
\newred{however, an important stabilizing mechanism in this region is
  found to be the $\mathbf{E}\times\mathbf{B}$ shear. Thus, a strong
  quench of the turbulence transport can be expected in this region.}
For the inner core radius, however, non-linear simulations show
destabilization of KBMs in the presence of toroidal rotation shear due
to the presence of a large PVG effect. The impact of PVG on the
destabilization of ITG modes has been reported previously in
Refs. \cite{kinseypop2005,barnsPRL2011,jonathanPRL2013}, however,
their impact on the destabilization of KBMs is reported here for the
first time, and further studies of this effect is ongoing.

In summary, a good inner core confinement due to strong stabilization
of the micro-turbulence driven transport can be expected in the hybrid
plasmas as a result of stabilization by the fast ion pressure which is
more effective at low magnetic shear due to $q$-profile optimization
achieved in the hybrid plasmas. For the outer core radii the hybrid
plasmas may benefit from a strong reduction of turbulent
  transport by $\mathbf{E}\times\mathbf{B}$ rotational shear, as was
reported in Ref. \cite{Litaudon2013,Irina}.

\section*{Acknowledgments}
The authors would like to thank J Candy for providing the \gyro
code. This project has received funding from the European Union's Horizon
2020 research and innovation programme under grant agreement number
633053. The views and opinions expressed herein do not necessarily
reflect those of the European Commission.  Computational resources of the High Performance
Computing for Fusion (HP-FF) facility were used for nonlinear
gyrokinetic simulations. IP is grateful for the financial support of
Vetenskapsr{\aa}det.

\end{document}